\documentclass{stsci_report}

\usepackage{multirow}
\usepackage{array}
\usepackage{graphicx}
\usepackage{siunitx}
\usepackage{todonotes}
\usepackage{pdfpages}
\usepackage{longtable}
\usepackage{booktabs}
\usepackage{float}
\usepackage{longtable}
\usepackage{indentfirst}
\usepackage{caption}
\usepackage{hyperref}
\usepackage{multirow}
\usepackage{paracol}
\usepackage{colortbl}
\usepackage{setspace} 
\usepackage{amsmath}
\usepackage{enumitem}
\usepackage{soul}
\usepackage{xcolor}
\usepackage{hyperref}
\hypersetup{
    colorlinks=true,
    linkcolor=black,
    urlcolor=blue,  
    citecolor=black
}
\usepackage{amssymb}
\newcommand{\farcs}{\mbox{$.\!\!^{\prime\prime}$}}

\copyrighttext{Copyright\copyright\ \the\year\ The Association of Universities for Research in Astronomy, Inc. All Rights Reserved.}

\presubtitle{Instrument Science Report WFC3 2025-05}
\title{Improved UVIS Aperture Corrections derived from Focus Diverse PSF Maps}
\author{K. Huynh, V. Bajaj, J. Mack, A. Calamida}
\date{August 18, 2025}

\begin{document}

\maketitle

\abstract{In crowded fields, small-aperture photometry can reduce contamination errors from neighboring sources compared to larger aperture photometry. However, the UVIS encircled energy (EE) varies with detector position and focus variations on orbital timescales for aperture radii less than 10 pixels ($\sim$0\farcs4). Using a set of focus-diverse empirical PSFs by \textcite{18Anderson}, we compute 2D spatial maps of the aperture correction between 5--10 pixels and find a maximum change of $\sim$0.01 mag over all focus levels for a given detector position. The upper-left and lower-right corners of the UVIS detector are more focus-sensitive than the rest of the field of view, where the mean correction is systematically $\sim$0.01 mag higher in Amp A for bluer filters (F275W, F336W, F438W) and $\sim$0.01 mag higher in Amp D for redder filters (F606W, F814W) at all focus levels. We test the new aperture correction maps in globular clusters, and we find reduced scatter, better agreement between the two CCDs, and a small shift in the absolute photometry when compared to a single (constant) aperture correction per image. These improvements are specific to photometry with apertures $<$ 10 pixels in radius; results from larger apertures are not affected. Using published EE tables can introduce systematic uncertainties in absolute photometry due to its tendency to vary with detector position and focus level, with larger errors for smaller apertures. Users requiring photometric accuracy better than $\sim$1\% for small apertures can use isolated stars in the individual \texttt{FLT}/\texttt{FLC} frames (or PSF cutouts at a similar detector position and focus level) to compute encircled energy corrections and accurately account for the amount of flux at radii larger than their photometric apertures.}

\newpage
\begin{spacing}{0.95}
    \tableofcontents
\end{spacing}

\newpage
\section{Introduction}
The shape and size of an observed point spread function (PSF) changes on short timescales due to ``breathing" effects, i.e. changes in the focal length caused by variations in the telescope's thermal environment such as going in and out of the Earth's shadow and changes in orientation relative to the Sun \parencite{Dressel2012}. PSFs also change shape over the Field of View (FoV) of the Wide Field Camera 3 Ultraviolet and Visible (WFC3/UVIS) channel due to the differences in material across the detector and changes in the geometry of the optics with location in the focal plane \parencite{18Anderson}. The WFC3/UVIS zeropoints are measured using 10 pixel aperture photometry, where variations in the PSF (and the encircled energy) from temporal and spatial variations are negligible ($\sim$0.1\%).
\bigbreak
However, when performing photometry in crowded stellar fields, a 10 pixel radius aperture can introduce contamination errors from neighboring sources when the wings of their PSFs overlap. While smaller apertures can mitigate these systematic errors \parencite{Dauphin2021}, they are more sensitive to temporal and spatial variations, particularly for UVIS apertures with radii less than eight pixels where substantial flux lies at larger radii \parencite{Sabbi2013}. Changes in the detector thickness and the tilt of the camera with respect to the focal plane are the major causes of variation of the full-width at half maxima (FWHM) of stars across the UVIS FoV.

\subsection{Focus-Diverse PSFs and ``Phylograms"}
\begin{figure}[ht]
\begin{center}
\includegraphics[width=\textwidth]{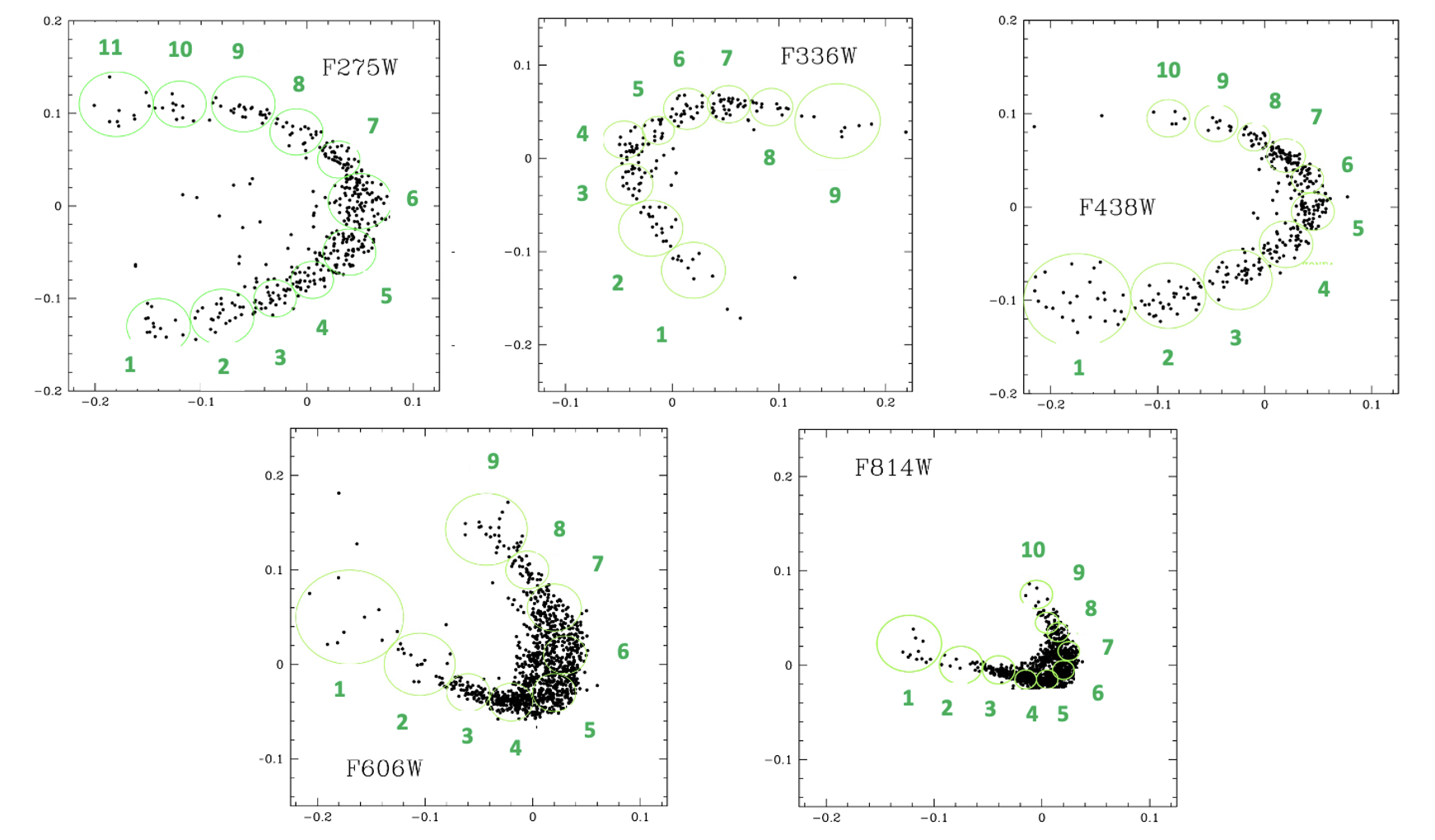}
\end{center}
\vspace{-1em}
\caption{\textit{Adapted from \textcite{18Anderson}, phylogram plots for F275W, F336W, F438W, F606W, and F814W, along with the phylo (focus) zones identified for each filter. The axes for all filters are arbitrary. The shape of the phylogram and the number of zones varies per filter. In all filters, the nominal phylo level (at best focus where the PSF is the sharpest) is located between levels 5--7.
}}
\label{fig:phylogram}
\end{figure}

\textcite{18Anderson} constructed focus-diverse PSF models from UVIS images of moderately-dense stellar fields for the five most-frequently used WFC3/UVIS filters: F275W, F336W, F438W, F606W, and F814W. Empirical measurements of the PSF asymmetry were computed for each exposure, and the PSF models were then grouped together based on their similarity. This leads to a natural ordering of the PSFs in a single curve shown in a two-dimensional (2D) grid, or a ``phylogram" plot, for each filter as shown in Figure \ref{fig:phylogram}.
\bigbreak
Each empirical focus level, or ``phylo" level, represents the PSF asymmetry, describing how sharp a PSF is due to the breathing of HST, with one side of the focus having an elongation in the $45^\circ$ direction and a $135^\circ$ elongation in the other. The phylo ranges are arbitrary, and the number of phylo zones and the shape of the phylogram varies per filter, with PSFs measured in redder filters being less sensitive to breathing. In all filters, the nominal phylo level where the PSF is the sharpest is located between phylo levels of 5--7 \parencite{18Anderson}. Using the \texttt{hst1pass} software \parencite{Anderson2022}, in conjunction with these focus-diverse PSF models, one could determine the phylo level of an individual exposure.
\bigbreak
Encircled energy (EE) refers to the fraction of flux contained within some radius relative to the theoretical total that reaches to infinite radius due to the PSF. To account for temporal and spatial variations in the EE, we develop a new method to compute 2D aperture correction maps that quantify variations in the PSF over the UVIS FoV. These maps are separated by the empirical focus level of PSFs measured in their respective filter and can be used to correct aperture photometry from a radius of 5--10 pixels to accurately account for the flux at larger radii.
\bigbreak
In this report, we present the aperture correction maps for five filters with focus-diverse PSFs (F275W, F336W, F438W, F606W, and F814W) and the methodology in creating them. These maps provide corrections for flux scattered to larger radii in the wings of the PSFs as a function of different focus levels, enabling accurate small-aperture photometry regardless of telescope breathing. We test improvements in aperture photometry at a radius of 5 pixels in two globular clusters with differing source densities: 1.) an uncrowded region 6 arcmin west of the core of NGC-104 (hereafter, 47 Tucanae) and 2.) the crowded core of the globular cluster Omega Centauri. We compare our results with the aperture correction maps to two standard methods: 1.) a constant aperture correction computed per image from the sigma-clipped mean of stars between a radius of 5--10 pixels, and 2.) aperture corrections derived from the published UVIS EE tables by \textcite{2021Calamida}.


\section{Methodology}
We create 2D aperture correction maps at each phylo level for the five filters with focus-diverse PSFs (F275W, F336W, F438W, F606W, and F814W). Our methodology is summarized as follows:
\begin{enumerate}
    \item All UVIS external science images (\texttt{FLC}) in the five designated wide-band filters are collected from the Mikulski Archive for Space Telescopes (MAST). 
    \item Using the \texttt{hst1pass} software \parencite{Anderson2022}, point-like sources in the images are detected and measured via PSF fitting, using the focus-diverse PSF models. The detector (X,Y) positions, fluxes, and quality of fits for all measured sources are recorded. Only images containing between 20--1500 sources are kept to mitigate sparse and intensely crowded field errors. Additionally, the phylo level is outputted by \texttt{hst1pass} based on measured PSFs in the exposure and is recorded for each image.
    \item Sources with a \texttt{hst1pass} quality of fit value $0 < q < 0.2$ are kept, leaving only well-measured sources.
    \item The remaining sources are then remeasured using standard \verb|photutils| aperture photometry.  The photometric aperture radius was set to 5 pixels, with a sky value calculated via a 3-$\sigma$ clipped mean measured in an annulus from 15--20 pixels, and subtracted from the photometric aperture.  This was then repeated with a 10 pixel photometric aperture and the same sky parameters.  The aperture correction for each star is calculated as $mag_{5 pix} - mag_{10 pix}$.
    \item We sort all sources by their filter and integer phylo value assigned to them based on their respective exposure. For each phylo group, we then calculate binned 2D statistics using every star's detector (X,Y) position coordinates and aperture correction as follows: We first group stars by their position on the detector into 512$\times$512 pixel bins covering the entire field. We then calculate the 2.5-$\sigma$ sigma-clipped median aperture correction from the grouped star aperture corrections computed in the previous step for every bin. Each 512$\times$512 pixel bin correspond to a single box in a 8$\times$8 grid. This produces a 2D map for the given filter, which provides an aperture correction at any coordinate for each phylo group. This step is then repeated for the four other filters, for a total of five maps, one for each filter.
    \item For any position bins within the maps that has an aperture correction 2-$\sigma$ different than the bins adjacent to it, We apply a Gaussian smoothing with a kernel defined by a standard deviation of 1 box width (i.e., 1 $\sigma$ = 1 box = 512 pixels). Additional smoothing is done to the Amp D corners of F275W and F438W maps to smooth over significant variations, which are noise introduced by the smaller number of sources observed near that corner of the CCD.
    
\end{enumerate}

When using these aperture correction maps, matching the exposure with the correct phylo number will allow the 5 pixel aperture photometry to be corrected for the temporal variation caused by breathing. The large sample of PSFs at different parts of the detector allow for sufficient spatial resolution across the detector to correct for the spatial variation. 

\section{Results}
\begin{figure}[H]
\begin{center}
\includegraphics[width=\textwidth]{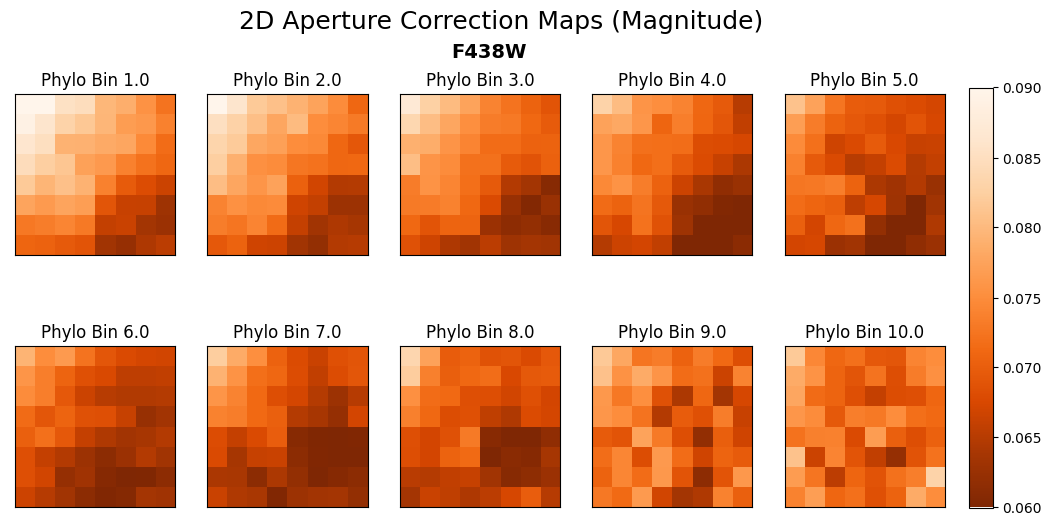}
\end{center}
\vspace{-1em}
\caption{\textit{2D spatial maps of the UVIS F438W aperture correction from 5--10 pixels (in magnitudes) derived from PSFs at different phylo (focus) levels. The aperture correction is greatest in the upper-left corner, where Amp A is located. The corrections are the smallest and the most flat across the detector at nominal phylo levels (5--7).
}}
\label{fig:2dmap438}
\end{figure}

\begin{figure}[H]
\begin{center}
\includegraphics[width=\textwidth]{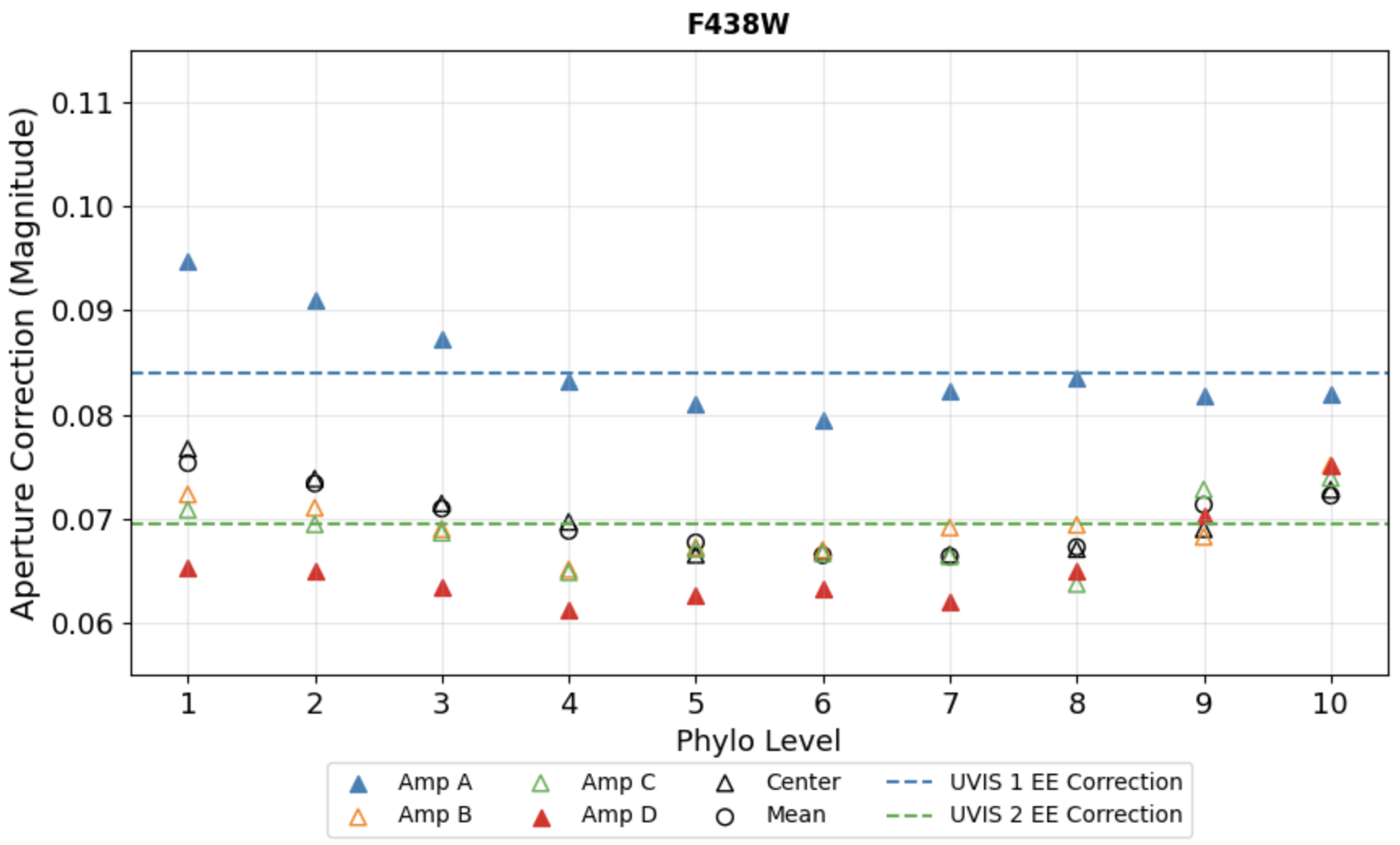}
\end{center}
\vspace{-1em}
\caption{\textit{F438W aperture correction (r = 5--10 pixels) versus phylo level for different regions of the UVIS detector. Colored triangles show a 512$\times$512 pixel cutout at the corner of amplifiers A, B, C, and D, corresponding to subarray apertures 'C512$\ast$-SUB'. Black triangles show the mean at the center of the UVIS FoV in a 1024$\times$1024 pixel cutout, while black circles show the mean over the entire FoV. Dashed lines are plotted for reference only, to compare with the aperture corrections from the published EE tables for UVIS1, Amp A at 0.085 mag (blue) and for UVIS2, Amp C at 0.070 mag (green). For F438W, the aperture correction for UVIS1 (UVIS2) is approximately equal to the mean of the blue (green) triangles over all phylo levels.}}
\label{fig:apcorrscatter438}
\end{figure}

\begin{figure}[H]
\begin{center}
\includegraphics[width=.9\textwidth]{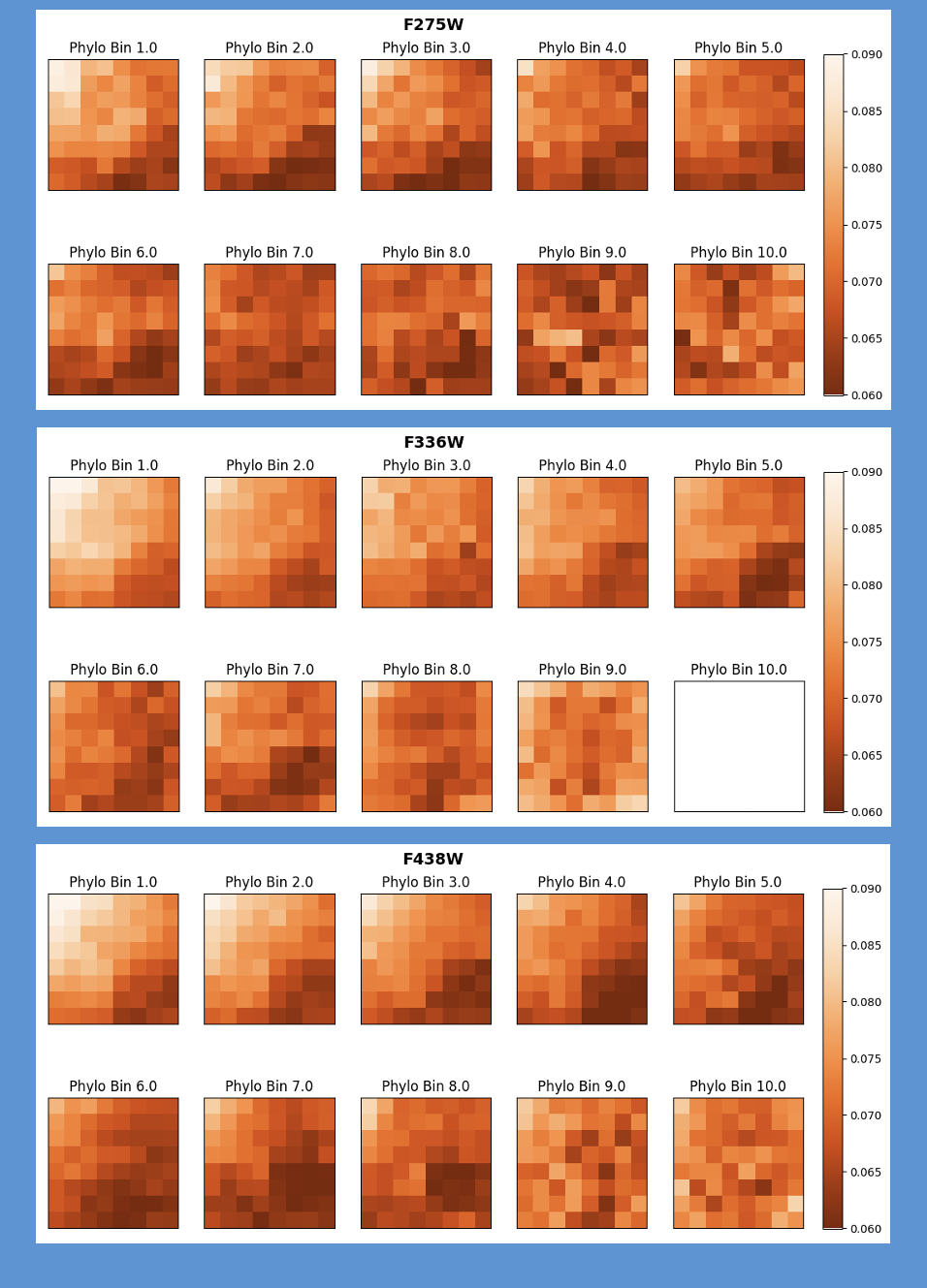}
\end{center}
\vspace{-1em}
\caption{\textit{2D spatial maps of the UVIS aperture correction from 5--10 pixels (in magnitudes) derived from focus-diverse PSF models at different phylo levels for bluer filters: F275W, F336W, and F438W. F336W was divided into only 9 focus groups, so phylo 10 map is blank for this filter.
}}
\label{fig:bluefilts}
\end{figure}

\begin{figure}[H]
\begin{center}
\includegraphics[width=.9\textwidth]{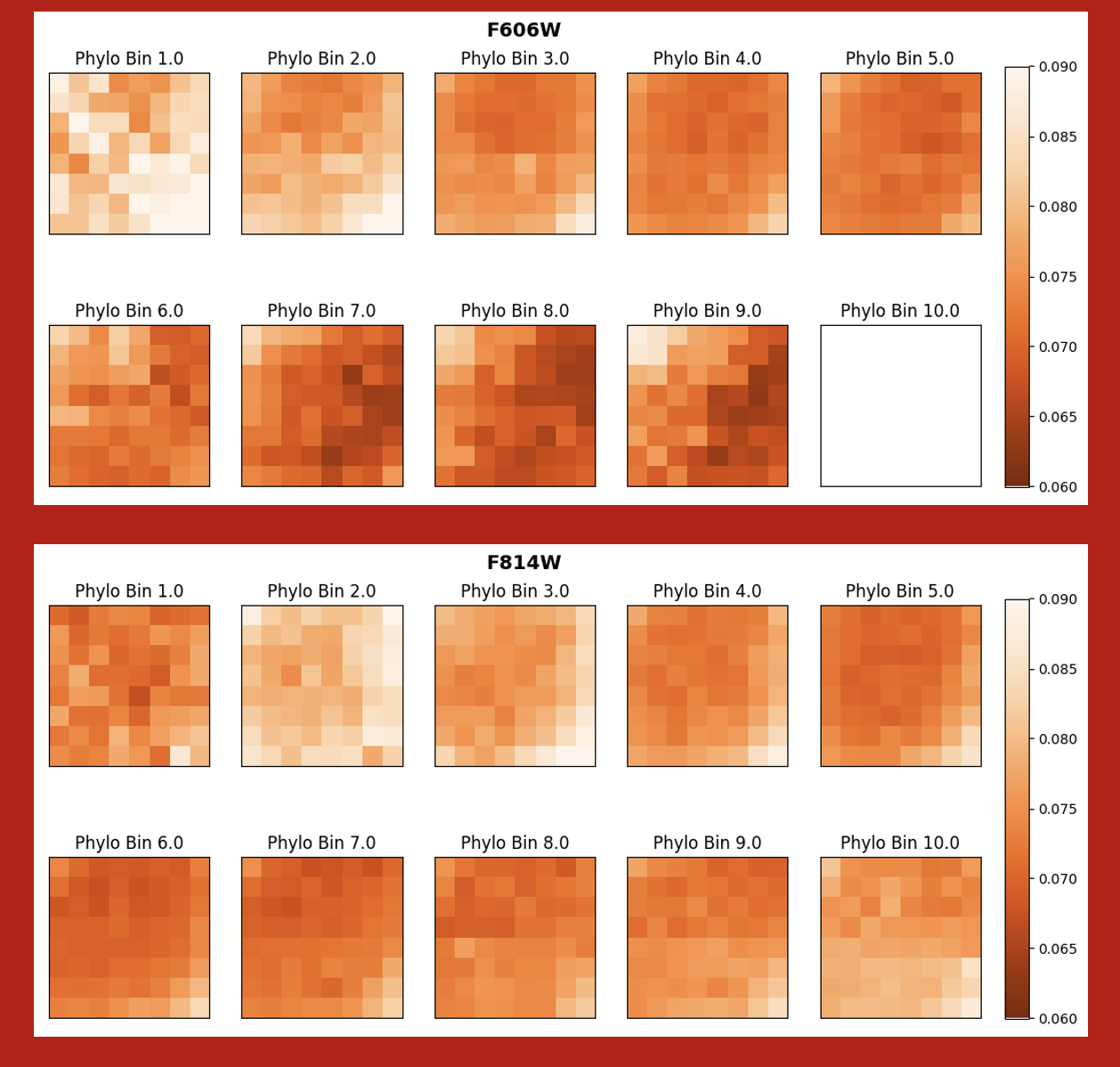}
\end{center}
\vspace{-1em}
\caption{\textit{Same as Figure \ref{fig:bluefilts}, except for redder filters: F606W and F814W. F606W was divided into only 9 focus groups, so phylo 10 map is blank for this filter.
}}
\label{fig:redfilts}
\end{figure}

\begin{figure}[H]
\begin{center}
\includegraphics[width=\textwidth]{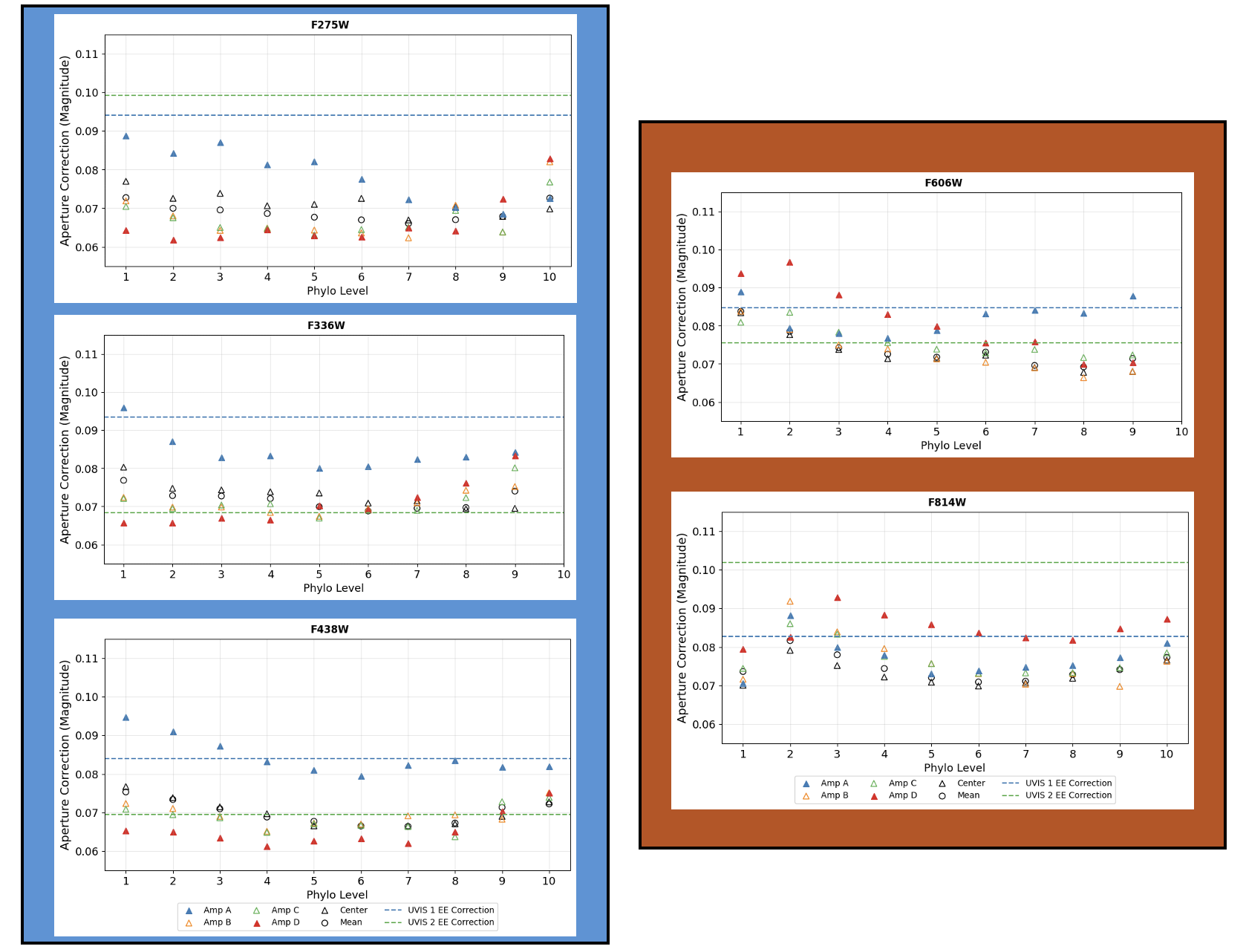}
\end{center}
\vspace{-1em}
\caption{\textit{Aperture correction (r = 5--10 pixels) versus phylo level for different regions of the UVIS detector (see Figure \ref{fig:apcorrscatter438}) for five filters: F275W, F336W, F438W, F606W, and F814W. Scatter plots are organized by color with bluer filters (F275W, F336W, and F438W) on the left and redder filters (F606W and F814W) on the right. Dashed lines are plotted to compare with aperture corrections calculated using the EE tables for UVIS1 (blue) and UVIS2 (green).  Aperture corrections are smaller at nominal phylo levels $\sim$5--7 where the telescope is in focus. This is seen in all filters in every region of the detector. The aperture correction in the Amp A corner (blue triangles) is $\sim$0.01 magnitudes higher than the rest of the FoV for most phylo levels in bluer filters. Conversely, the aperture correction in the Amp D corner (red triangles) is $\sim$0.01 magnitudes higher in redder filters for most phylo levels.
}}
\label{fig:apcorscatter_comb}
\end{figure}

\hypertarget{sec:results}{}
Figure \ref{fig:2dmap438} shows the focus and spatially dependent aperture correction map in units of magnitudes for phylo levels 1--10 in F438W. The maps cover the entire size of the FoV binned down to a 8$\times$8 grid. Additionally, Figure \ref{fig:apcorrscatter438} shows the aperture correction (in magnitudes) versus phylo levels for 512$\times$512 pixel cutouts at the amplifier corners of the detector (A, B, C, and D). Also shown are the mean aperture correction for a 1024$\times$1024 cutout at the center of the FoV, the mean aperture correction over the entire FoV, and dashed lines for the 5--10 pixel aperture correction using the published UVIS EE tables for both chips. 
\bigbreak
Figures \ref{fig:bluefilts} and \ref{fig:redfilts} shows the 2D aperture correction maps for blue (F275W, F336W, and F438) and red (F606W, and F814W) filters, respectively. Additionally, Figure \ref{fig:apcorscatter_comb} shows the aperture correction versus phylo levels for all five filters. Figure \ref{fig:apcorscatter_comb} is organized by wavelength, with bluer filters (F275W, F336W, F438W) and redder filters (F606W and F814W) together. There are only 9 phylo groups for F336W and F606W, hence the absence of a spatial map for phylo 10 in those filters.


For all filters, we find that the amplitude of the aperture corrections is smaller at nominal phylo levels (5--7) where the PSF is the sharpest.  Grouping the filters together by color, we notice similarities between the aperture corrections for different regions of the UVIS detector. In bluer filters, the Amp A corner is consistently higher than the rest of the FoV by $\sim$0.01 mag for most phylo levels. Similarly, in redder filters, the Amp D corner is $\sim$0.01 mag higher than the rest of the FoV for most phylo levels. This is consistent with the results from \textcite{Sabbi2013}, who shows that the observed PSFs are less well-focused near the A and D amplifiers due to the tilt of the UVIS detector with the respect to the focal plane.  

\bigbreak
There are also some distinct behavior in specific filters. In F275W, the Amp D corner exhibits a high aperture correction for phylo levels 9 and 10, even after smoothing out those specific corners. This may likely be due to small number of statistics, as only 5\% of all stars used to make the F275W maps have a phylo level between 9--10. This may also be attributed to the behavior of PSF sources at levels 9--10, where the PSFs are extremely out of focus and increases the scatter of the measured aperture corrections. Furthermore, at low phylo levels in F606W, the Amp D corner has a higher aperture correction with respect to the rest of the FoV, while the aperture correction in the Amp A corner is uniform with the detector. However, as we increase to higher phylo levels, we find that the correction for Amp D returns to uniform with the rest of the detector, and Amp A increases in magnitude, reversing the behavior between the two cross amps that is exhibited at lower phylo levels. The spatially variable aperture corrections are also consistent with the aperture corrections measured using the UVIS EE tables for every filter except F275W and F814W, where the UVIS2 encircled energy aperture correction is substantially higher than every amplifier corner.





\section{Aperture Correction Validation in 47 Tucanae’s Uncrowded Field}

We revisit two early WFC3 calibration (CAL) programs that observed an uncrowded region 6 arcmin west of the core of the galactic globular cluster 47 Tucanae. These consist of 350 seconds exposures with large dithers across the UVIS FoV. They are useful for testing the new aperture correction maps, as dithered exposures allow us to quantify any residual variations in the brightness of individual stars at different positions on the detector.
\bigbreak

\begin{figure}[H]
\begin{center}
\includegraphics[width=\textwidth]{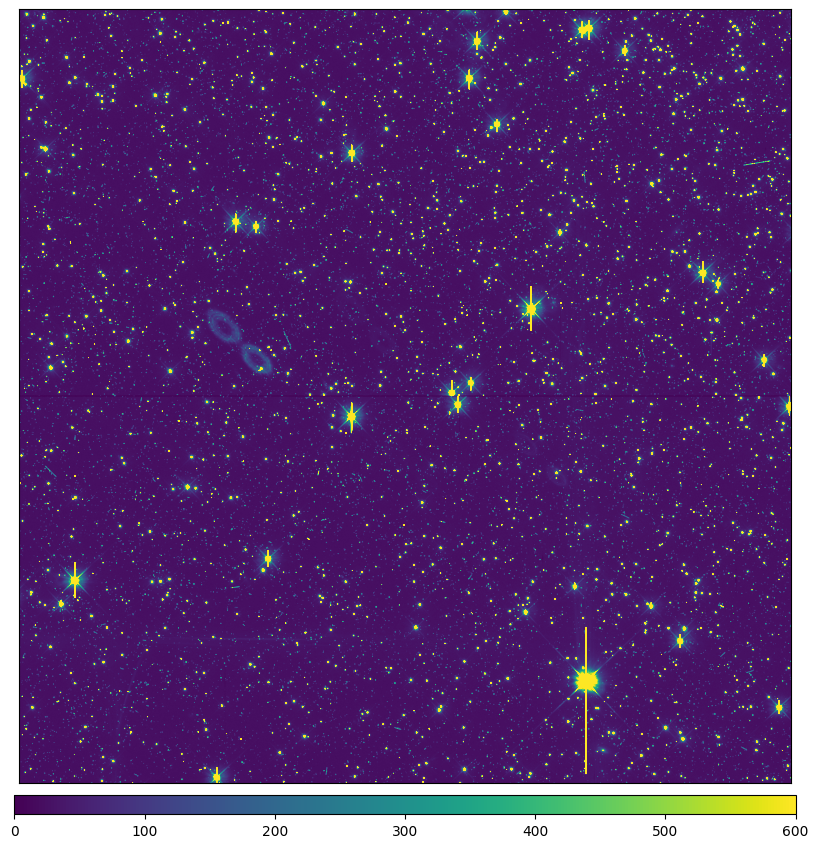}
\end{center}
\vspace{-1em}
\caption{\textit{Sample 350 second F606W exposure \texttt{iaby01kwq} showing an uncrowded region in 47 Tucanae, 6' west of the cluster core. Dithered exposures were acquired in program 11452 with nine pointings in a large 3$\times$3 box pattern, shifting by $\sim$25\% of the UVIS FoV. Additional exposures were acquired in program 12379 and consist of 2 dither positions offset in the Y-direction by $\sim $50\% of the UVIS FoV. The colorbar is in units of electrons.}}
\label{fig:47tucimg}
\end{figure}

CAL program 11452 was designed to measure the UVIS flat field uniformity across the detector in nine pointings in a large 3$\times$3 box pattern, with dither steps over $\sim$25\% of the FoV \parencite{Sabbi2009}.  While this program observed using six filters, we test only the F438W, F606W, and F814W filters and defer analysis of the UV filters to future work. CAL program 12379 revisited 47 Tucanae to investigate the effects of field crowding on the UVIS Charge Transfer Efficiency (CTE) \parencite{2024Kuhn}. From this program, we use pairs of F606W exposures dithered by  $\sim$50\% of the FoV to test the new aperture corrections. Table \ref{tab:47tucdata} in the \hyperlink{sec:appendix}{Appendix} lists the rootname, filter, proposal ID, and \texttt{hst1pass} phylo values for all 47 Tucanae exposures. Figure \ref{fig:47tucimg} shows a sample F606W exposure from CAL program 12379.
\bigbreak
We compare our phylo aperture correction maps against a constant (sigma-clipped mean) aperture correction derived per image, and the chip-dependent aperture corrections derived using the EE tables. A constant aperture correction per image accounts for temporal effects due to focus changes, however it does not account for spatial variations, which can be a larger effect.
\bigbreak
The UVIS EE tables were derived from images of isolated CALSPEC stars measured out to large radii (e.g. $\sim$6\farcs0 or 151 pixels) and were used to derive the UVIS photometric calibration \parencite{2021Calamida}. The EE tables provide a constant correction for each CCD chip in all 42 UVIS filters and were derived by combining many exposures with a wide range of focus levels. EE values were measured in the corners of Amps A and C, where the zeropoints are computed for each UVIS CCD in order to minimize losses from CTE. As shown in Figure \ref{fig:apcorscatter_comb}, the EE in the amplifier corners can be significantly different from the EE in the center of the detector for small apertures. UVIS EE tables can be downloaded on the WFC3 webpage \href{https://www.stsci.edu/hst/instrumentation/wfc3/data-analysis/photometric-calibration/uvis-encircled-energy}{here}, however please see the recommendations in Section \hyperlink{sec:rfu}{7}.

\bigbreak
We notice that aperture corrections derived from the EE tables are slightly larger than the phylo-based maps in Figure \ref{fig:apcorscatter_comb}, and we attribute some of this difference to the different sky annulus used for each method. While the EE tables measure the sky in an annulus between 160 and 200 pixels, the aperture correction maps are derived from shorter exposures and therefore utilize a smaller sky annulus between 15 and 20 pixels. This results in aperture corrections which are slightly smaller than predicted from the EE tables due to subtracting a small fraction of the PSF wings. To estimate the effect of the different sky annulus, we recalculate the UVIS EE using a smaller sky annulus from 15-20 pixels, and we find that this lowers the predicted aperture correction by $\sim$0.01\ mag in the F814W filter, for example. 
\bigbreak
In Figure \ref{fig:tuccompare438}, and Figures \ref{fig:tuccompare606} and \ref{fig:tuccompare814} in the Appendix, we compare the aperture corrections from different methods for two 47 Tucanae images per filter: one at nominal phylo and the other at an extreme phylo for three UVIS filters: F438W, F606W, and F814W, respectively. In all cases, the constant aperture correction is consistent with the average phylo aperture correction for that image. Thus, in uncrowded fields, we verify that the constant aperture correction can be accurately measured and is similar to the phylo-based corrections. In contrast, the aperture corrections computed using EE tables are overestimated in most cases. When applied to the observed photometry, these corrections can systematically bias the results, introducing errors in absolute photometry.

\begin{figure}[H]
\begin{center}
\includegraphics[width=1.0\textwidth]{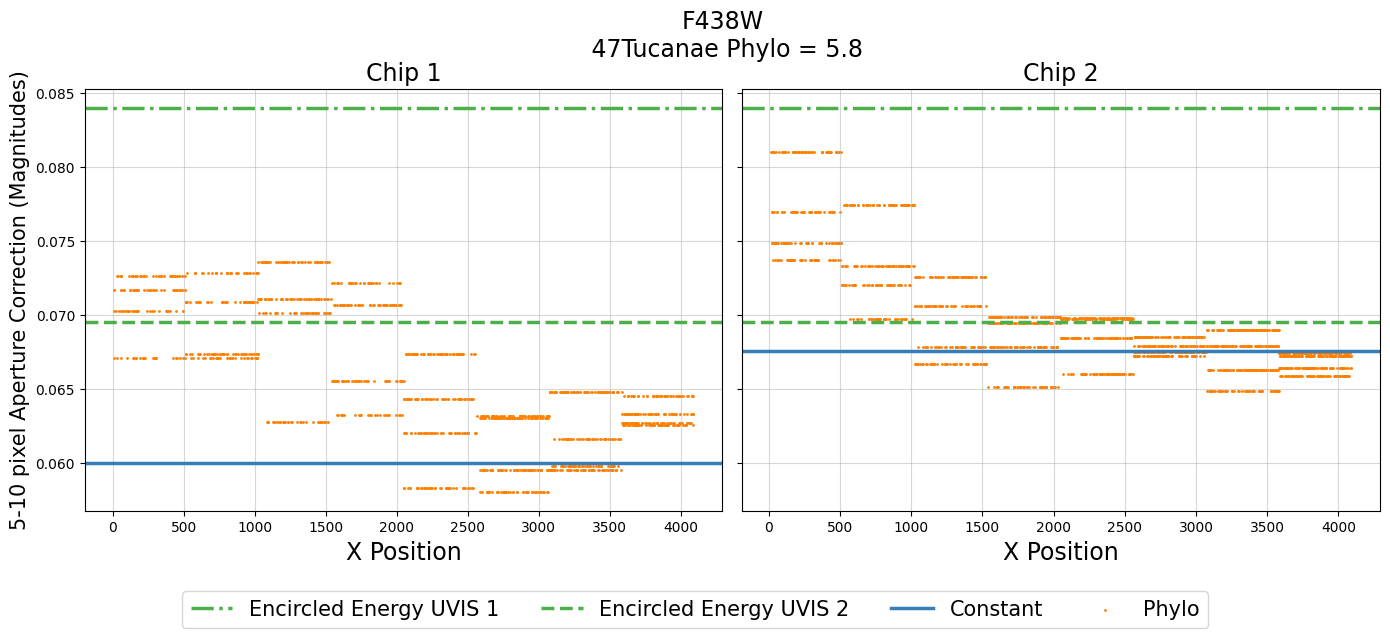}
\bigbreak
\includegraphics[width=1.0\textwidth]{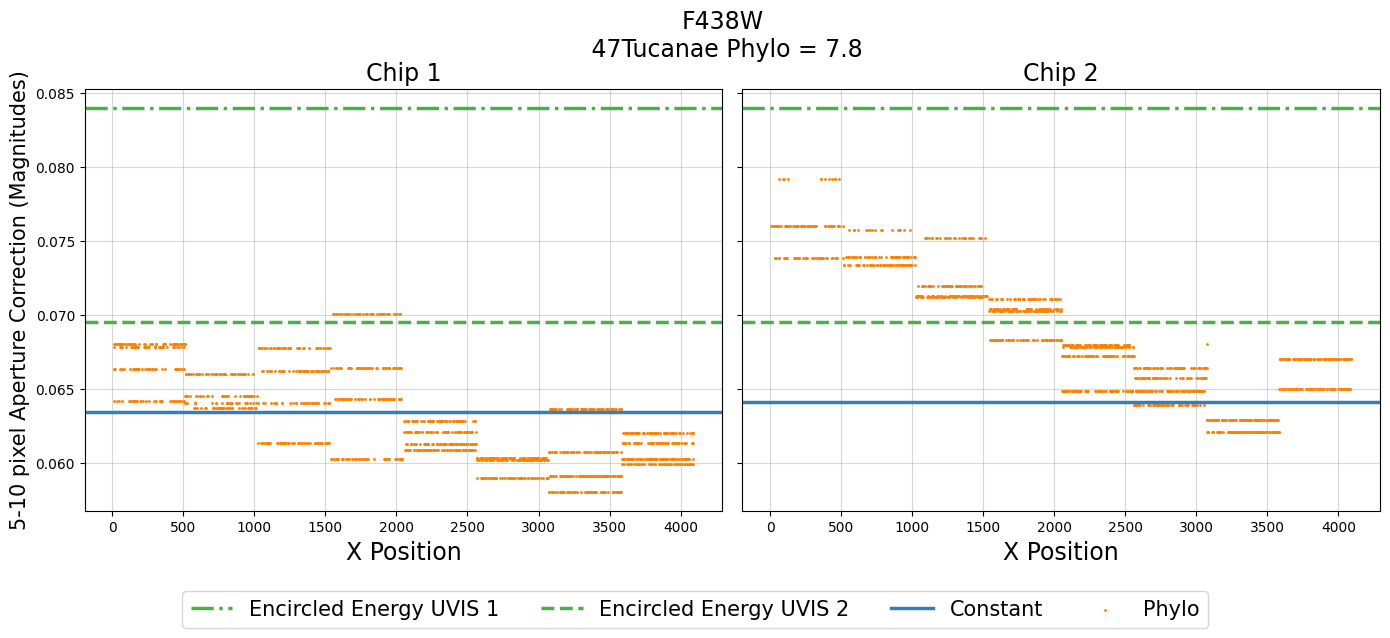}
\end{center}
\vspace{-1em}
\caption{\textit{F438W aperture correction (r = 5--10 pixels) versus the X-position of stars for two exposures in the uncrowded region of 47 Tucanae: one at nominal phylo = 5.8 (top) and another at an extreme phylo = 7.8 (bottom). The phylo-based aperture corrections for each star (orange) are consistent with the constant aperture correction (blue) and confirms that the constant aperture correction can be reliably measured in an uncrowded field. For reference, aperture corrections derived from the encircled energy tables are overplotted in green, where the dashed line for UVIS2 (Amp C) is generally consistent with the rest of the UVIS FoV and the dot-dashed line for UVIS1 (Amp A) is $\sim$0.01 mag higher.}
}
\label{fig:tuccompare438}
\end{figure}

\subsection{Color-Magnitude Diagrams from Different Aperture Correction Methods}
\begin{figure}[H]
\begin{center}
\includegraphics[width=\linewidth]{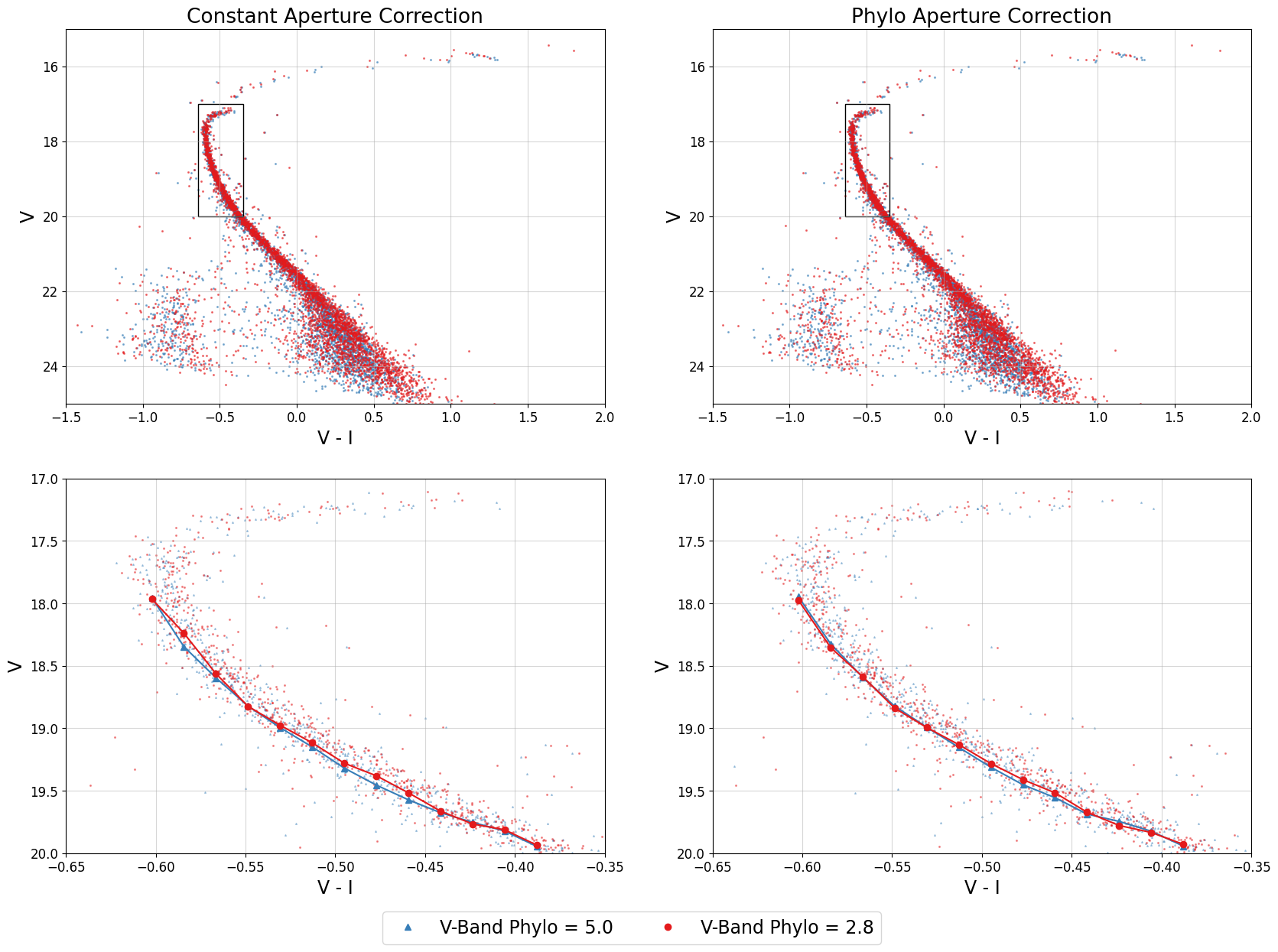}
\end{center}
\vspace{-1em}
\caption{\textit{47 Tucanae color-magnitude diagrams derived from WFC3/UVIS V- and I-band (F606W and F814W) photometry in a 5 pixel aperture, corrected to 10 pixels using a constant (sigma-clipped mean) aperture correction per image (left) versus phylo-based aperture corrections (right). We use two V-band exposures, one at nominal phylo = 5.0 (\texttt{ibnh11bmq}, in blue) and another at extreme phylo = 2.8 (\texttt{ibnh14x7q}, in red). In order to minimize the number of free parameters, the I-band magnitudes represent the mean magnitude of matched stars measured from all F814W images. Bottom plots are zoomed-in cutouts of the cluster main-sequence. The median V-band difference between the blue and red points is 0.016 magnitudes across the entire main-sequence cutout when using a constant aperture correction per image for exposures of differing phylos. Conversely, the median V-band difference is reduced to 0.002 magnitudes when using the phylo-based aperture corrections.}
}
\label{fig:47tuccmd}
\vspace{-1em}
\end{figure}

Color-magnitude diagrams (CMDs) derived from multiple exposures in the same filter with similar exposure times will inherit a shift in the observed photometric measurements per star due to the spatial and temporal effects. After correcting for these effects using our 2D aperture correction maps, the dispersion in the color-magnitude diagrams from exposures of differing phylo levels should decrease, compared to the CMDs derived using a constant aperture correction per exposure.
\bigbreak
Figure \ref{fig:47tuccmd} shows CMDs derived from V- and I-band (F606W and F814W) photometry in a 5 pixel aperture, corrected to 10 pixels using a constant aperture correction per image versus our phylo-based aperture correction. Each panel shows two different V-band exposures: \texttt{ibnh11bmq} at nominal phylo = 5.0 and \texttt{ibnh14x7q} at extreme phylo = 2.8. For the I-band photometry, we compute the mean magnitudes of matched stars measured from all 47 Tucanae images in the F814W dataset in order to minimize the number of free parameters.
\bigbreak
We plot the median-binned statistic across a main-sequence cutout containing bright and unsaturated stars. When using the constant aperture correction, we find a mean shift of 0.016 magnitudes in the V-band magnitudes between the extreme and nominal phylo exposures for stars located among a main-sequence cutout. Conversely, the V-band magnitude differs by only 0.002 magnitudes across the same region when using the phylo aperture correction maps. This verifies that the new corrections improve the photometric repeatability between exposures acquired at different phylo (focus) levels.

\newpage
\subsection{Magnitude Residuals for Stars Dithered over the FoV}
\begin{figure}[H]
\begin{center}
\includegraphics[width=\textwidth, trim={5em, 2em, 5em, 2em}, clip]{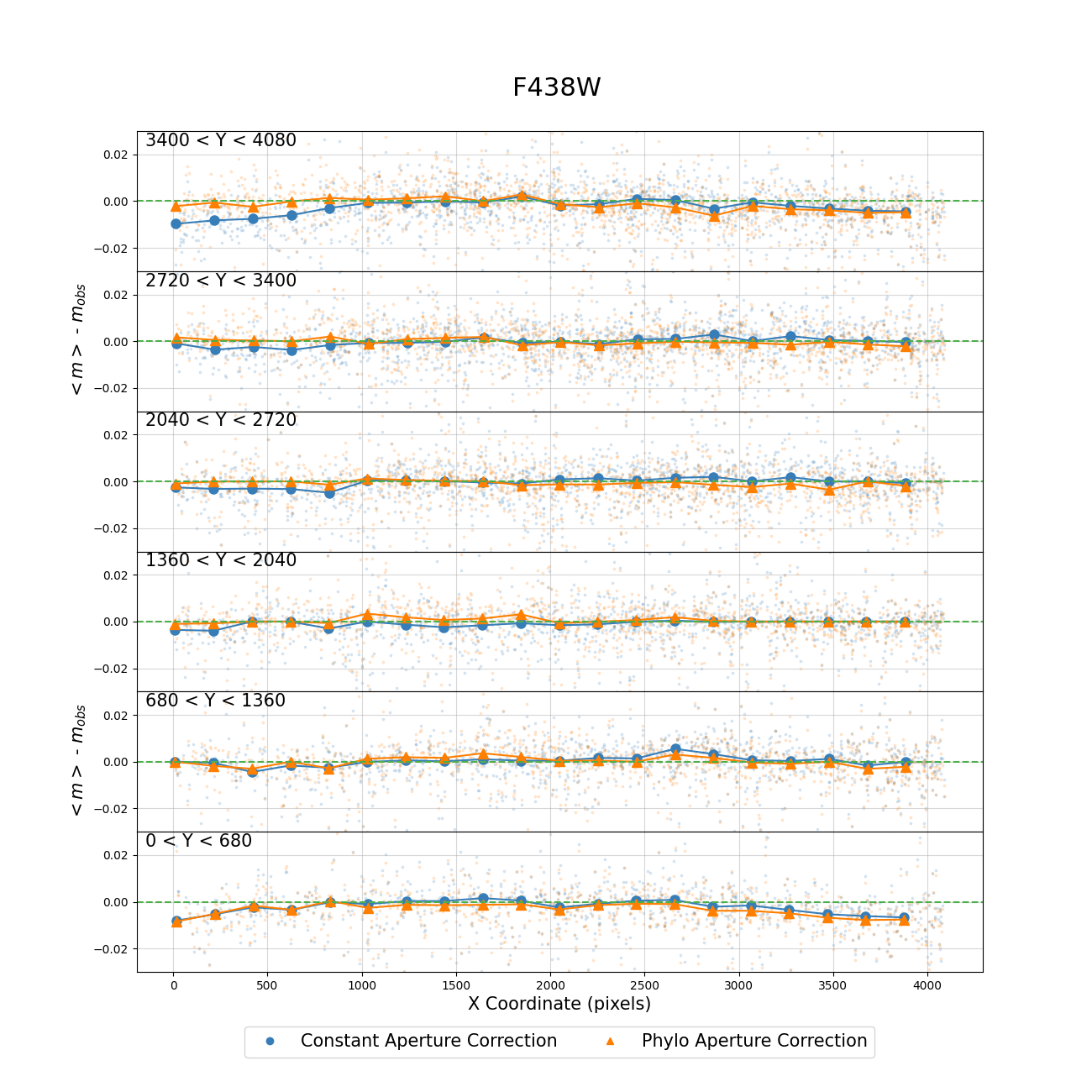}
\end{center}
\caption{\textit{F438W magnitude residuals (mean magnitude - individual magnitude of each star) versus X-position for stars in 47 Tucanae measured in six regions of the detector separated by Y-position. The fainter points show the individual magnitude residuals. Solid points show the median residual using $n=$ 20 bins to better visualize differences in photometry when using phylo-based corrections (orange) and constant aperture corrections (blue). For reference, a dashed green line is drawn where the difference is 0. The phylo-based corrections improve the spatial residuals, especially in the corner of Amp A (upper-left), where the improvement is $\sim$0.01 magnitude compared to a constant aperture correction.
}}
\label{fig:magres438}
\end{figure}

\begin{figure}[H]
\begin{center}
\includegraphics[width=\textwidth, trim={5em, 2em, 5em, 2em}, clip]{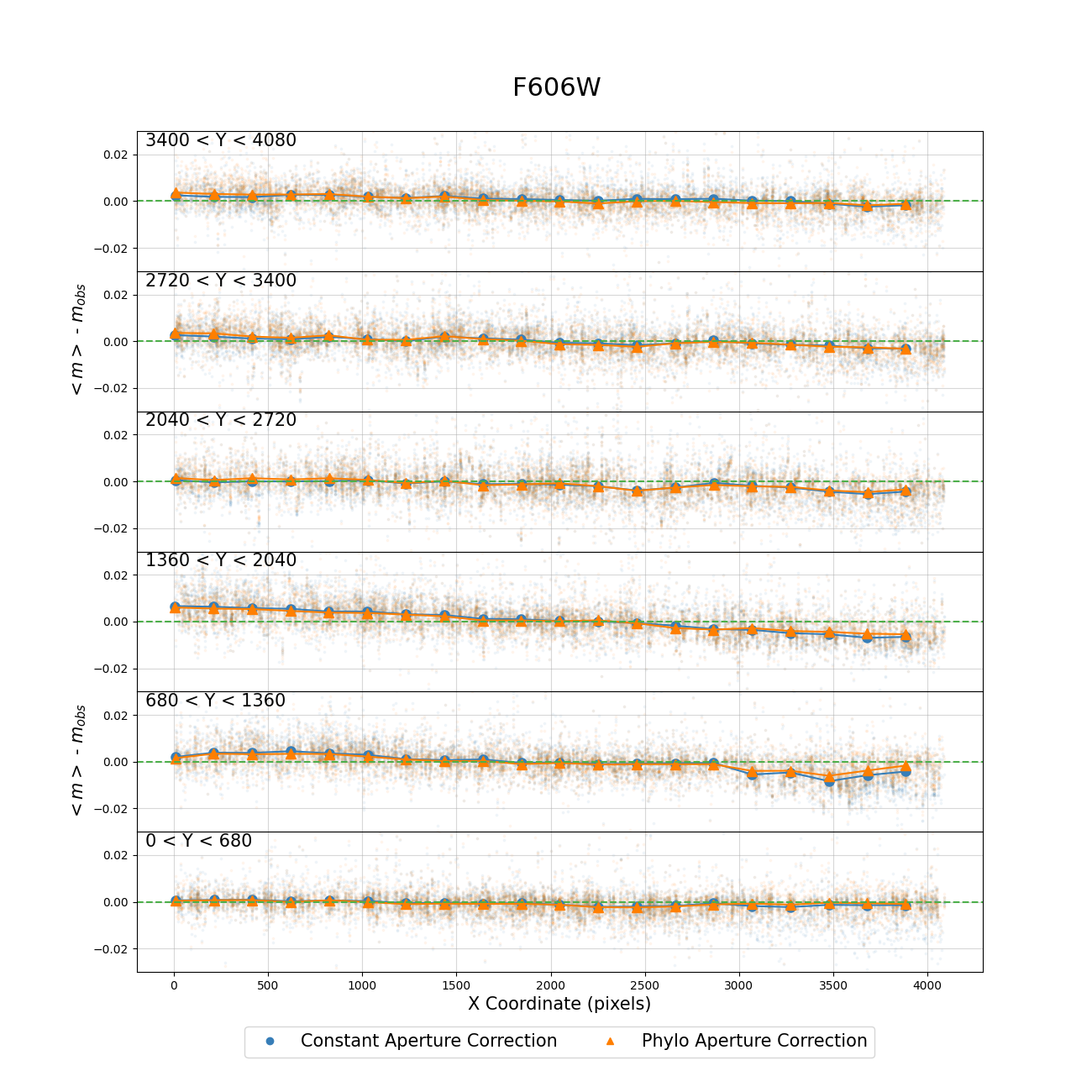}
\end{center}
\caption{\textit{Same as Figure \ref{fig:magres438}, but for F606W. The magnitude residuals between the two aperture correction methods are similar across the entire FoV, apart from a small offset in Amp D (lower-right), where the phylo-based aperture correction slightly improves the magnitude residual.
}}
\label{fig:magres606}
\end{figure}

\begin{figure}[H]
\begin{center}
\includegraphics[width=\textwidth, trim={5em, 2em, 5em, 2em}, clip]{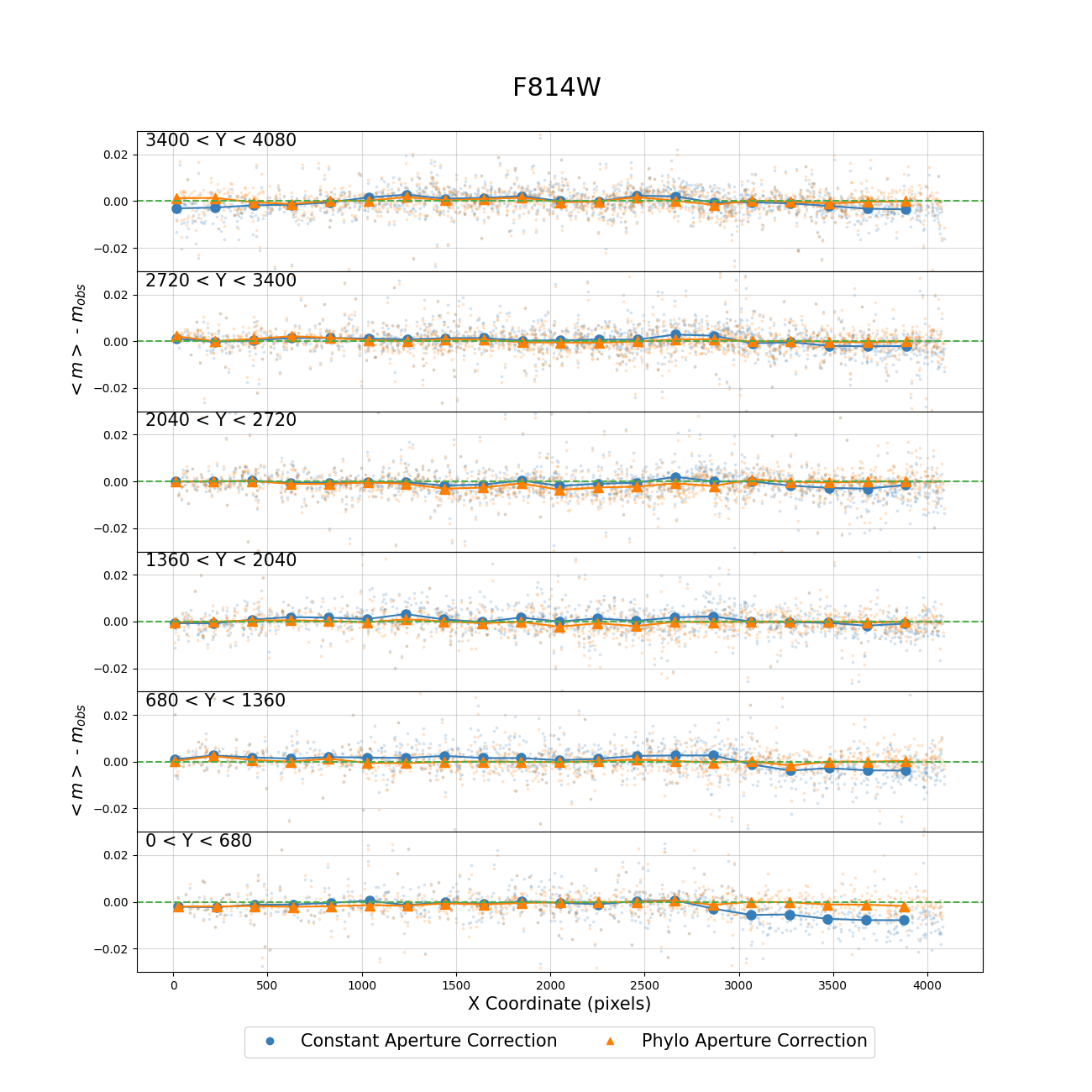}
\end{center}
\caption{\textit{Same as Figure \ref{fig:magres438}, but for F814W. The phylo-based aperture corrections improve the spatial residuals, especially in the corner of Amp D (lower-right), where the improvement is $\sim$0.01 magnitudes compared to a constant aperture correction.
}}
\label{fig:magres814}
\end{figure}

Ideally, measurements of the same non-variable star taken at different parts of the detector should give the same observed photometry, assuming a perfect flat field correction. In practice, the tilt of the UVIS detector in the focal plane may affect the measurement at different parts of the detector, leaving a residual when comparing photometry in small apertures. Our new aperture correction maps seek to reduce the magnitude residual of a given star, so that the residual of measurements across the detector is near zero. 
\bigbreak
To assess this improvement, we select bright, unsaturated stars in 47 Tucanae observed at different positions across the detector and computed the average magnitude per UVIS chip. We then took that average magnitude, subtracted each individual magnitude measurement, and plotted the photometric residuals as a function of the star's X-coordinate, in six different Y-coordinate bins spanning the entire UVIS FoV. Figures \ref{fig:magres438}, \ref{fig:magres606} and \ref{fig:magres814} show the magnitude residual versus star position for filters F438W, F606W and F814W, using a constant aperture correction versus our phylo aperture correction. Medians in 20 X-coordinate bins across the detector  are over plotted in order to better visualize the magnitude residual differences.
\bigbreak
The residual offset between the phylo and constant aperture corrections for F438W are relatively uniform throughout the FoV, except in the Amp A (top left) corner where the phylo aperture correction reduces the magnitude residual by $\sim$0.01 mag compared to a constant correction. In F606W, the entire FoV have similar magnitude residuals between the two corrections, with a very small offset on the right edge of the UVIS1 detector, where the phylo aperture correction improves the magnitude residuals slightly. For F814W, we find a similar residual decrease of 0.01 mag, but in the Amp D corner (bottom right) instead of the Amp A corner for F438W.

\bigbreak
The improvement in the magnitude residuals agrees with the behavior of our spatially variable aperture correction maps. As mentioned in the \hyperlink{sec:results}{Results}, bluer filters like F438W have a $\sim$0.01 magnitude higher aperture correction in the Amp A corner at most focus levels, while redder filters like F814W have a $\sim$0.01 magnitude higher aperture correction in the Amp D corner.

\section{Aperture Correction Validation in Omega Centauri's Crowded Stellar Field}

\begin{figure}[H]
\begin{center}
\includegraphics[width=1.0\textwidth]{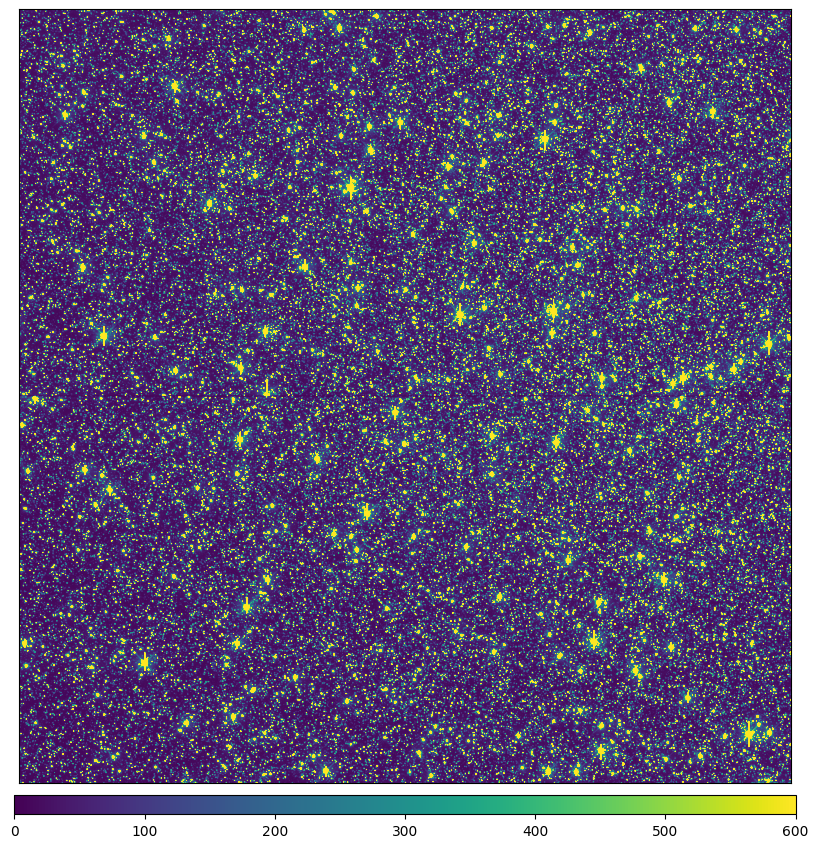}
\end{center}
\vspace{-2em}
\caption{\textit{Sample 40 second F606W exposure \texttt{ibc302ivq} showing the crowded core of Omega Centauri. Dithered exposures of this field were acquired in program 11911 with 9 pointings in a large 3$\times$3 box pattern, with dither steps shifting by $\sim$25\% of UVIS FoV. The crowding effects are much more apparent compared to the observed field in 47 Tucanae. The colorbar is in units of electrons.}
}
\label{fig:ocenimg}
\end{figure}

A constant aperture correction per image agrees well with the phylo-based aperture correction in uncrowded areas, however they are much more difficult to estimate when observing a crowded region due to contamination from background stars in the aperture annuli. We continue to validate the photometric improvements from our phylo-based aperture corrections by performing 5 pixel aperture photometry around sources in the crowded core of Omega Centauri across several dithered exposures in F438W, F606W, and F814W. We compare the results using the phylo-based aperture correction versus a 5--10 pixel sigma-clipped mean constant aperture correction estimated per image.
\bigbreak
Omega Centauri exposures are taken from CAL programs 11911 and 12339, which were used to develop the in-flight corrections to the UVIS flat fields across the ten most commonly used filters, including F438W, F606W, and F814W \parencite{2013mack}. Similar to 47 Tucanae observations from CAL proposal 11452, exposures were taken using nine pointings in a large 3$\times$3 box pattern, with dither steps of $\sim$25\% of the FoV in the X and Y direction. We only select stars that have no brighter neighbors within 5 pixels to further constrain possible contamination effects. Table \ref{tab:ocendata} in the \hyperlink{sec:appendix}{Appendix} presents the rootname, filter, proposal ID, and phylo value measured in all Omega Centauri exposures used in the validation of our aperture correction. A sample F606W exposure of Omega Centauri is shown in Figure \ref{fig:ocenimg}.

\begin{figure}[H]
\begin{center}
\includegraphics[width=\linewidth]{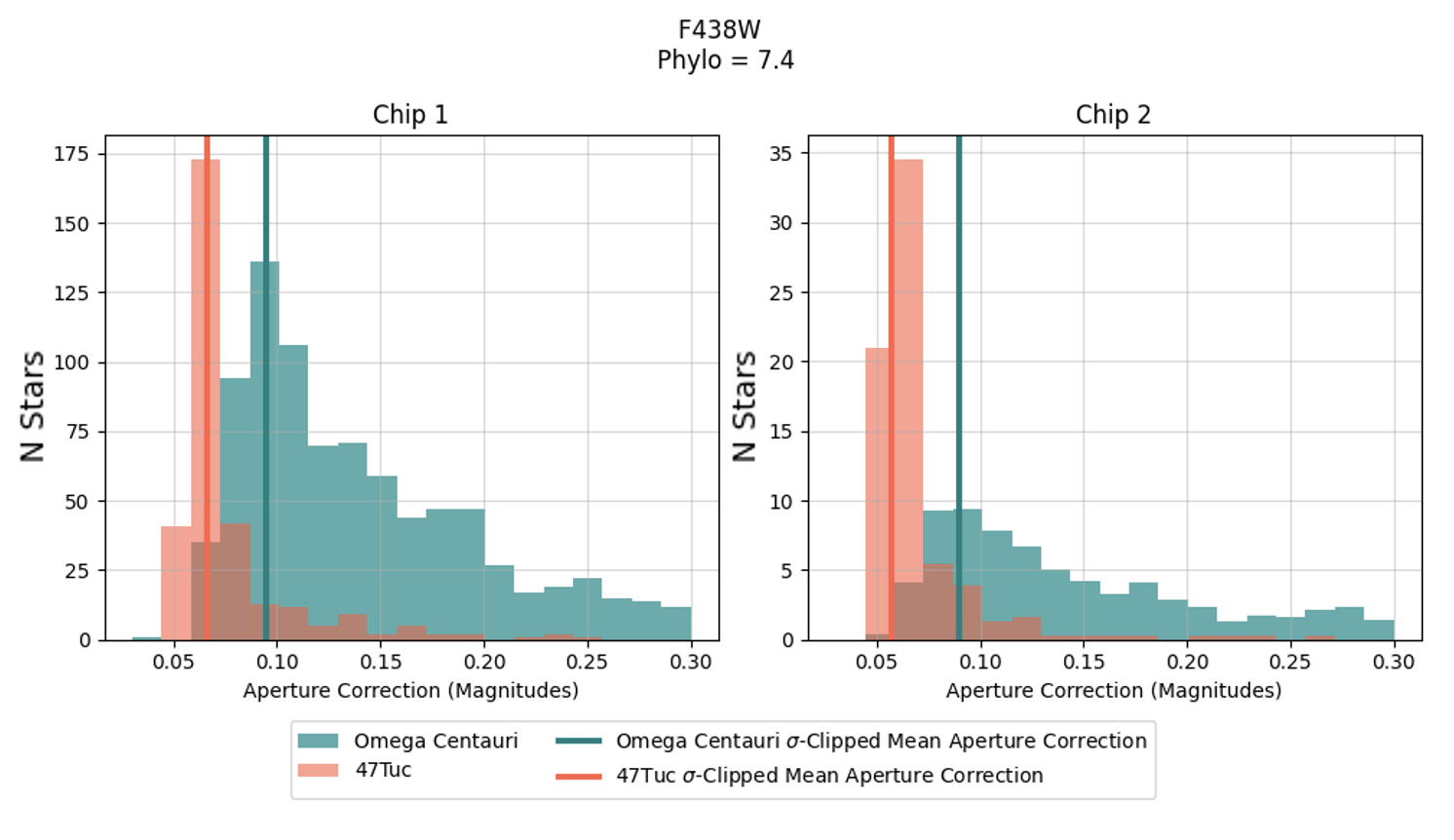}
\end{center}
\vspace{-2em}
\caption{\textit{Histograms showing the aperture correction for individual stars for two exposures at the same phylo level, comparing the crowded core of Omega Centauri (\texttt{ibc308xuq}) and the uncrowded region of 47 Tucanae (\texttt{iaby01lgq}). The constant aperture correction, computed from the $\sigma$-clipped mean of all stars, is overplotted. The scatter in the core of Omega Centauri is much higher than in 47 Tucanae, systematically shifting the computed aperture correction to a higher value. This illustrates the effects of contamination in crowded fields, where it can be more difficult to accurately measure aperture corrections from individual exposures.
}}
\label{fig:targhisto}
\vspace{-1em}
\end{figure}

Figure \ref{fig:targhisto} shows histograms of the aperture corrections in magnitudes measured for each individual star for two F438W images of the same phylo value = 7.4 for the crowded core of Omega Centauri and the uncrowded region of 47 Tucanae. The sigma-clipped constant aperture correction applied to the images are also displayed in the histogram for UVIS1 and UVIS2. The individual aperture corrections measured for stars in the crowded core of Omega Centauri are significantly more scattered compared to that of 47 Tucanae, which systematically shifts the sigma-clipped mean aperture correction applied to the image to be higher than estimated. This is due to the contamination effects of background sources in the apertures, demonstrating the difficulties of measuring an accurate constant aperture correction per image in a crowded field versus a non-crowded one.

\begin{figure}[H]
\begin{center}
\includegraphics[width=\linewidth]{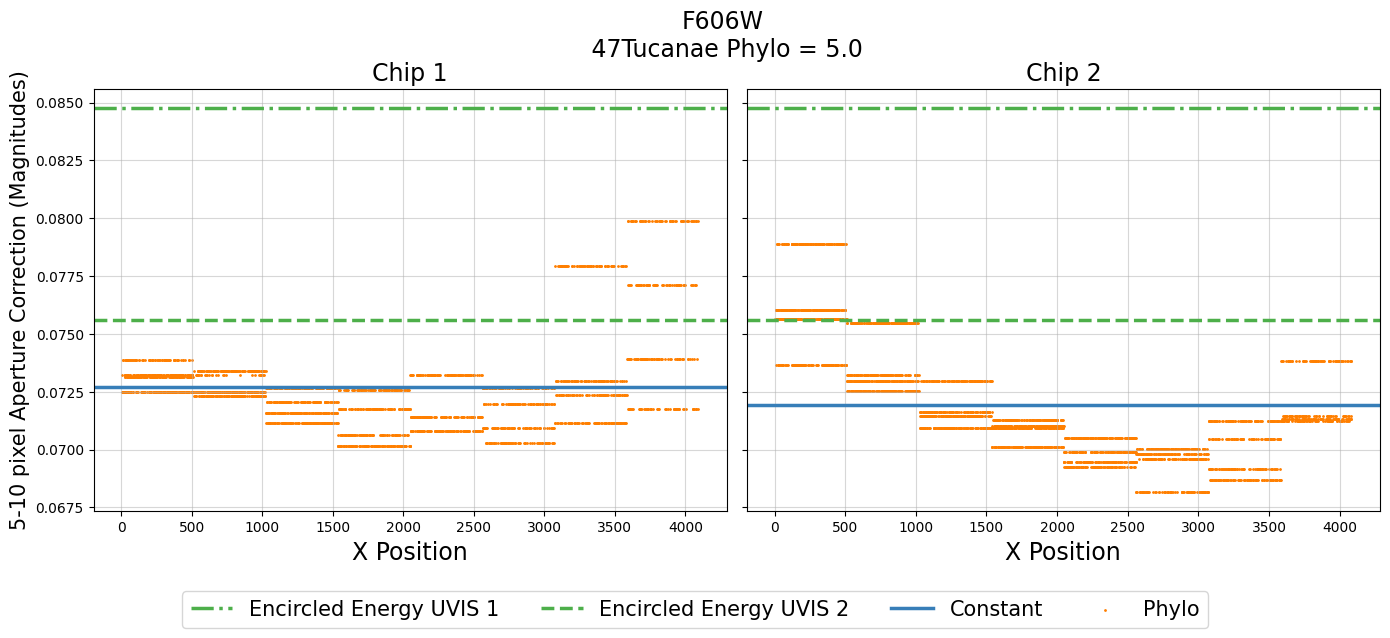}
\bigbreak
\includegraphics[width=\linewidth]{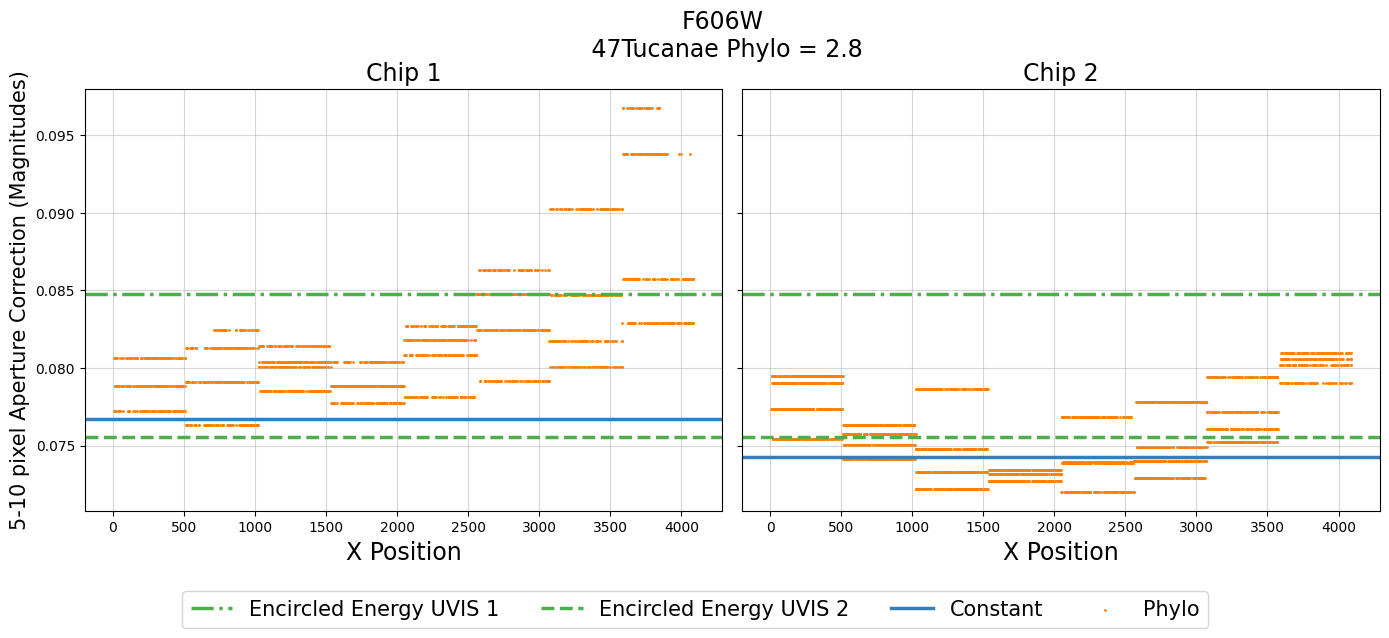}
\end{center}
\caption{\textit{Same as Figure \ref{fig:tuccompare438}, but for two F606W exposures of 47 Tucanae: one at nominal phylo = 5.0 (top) and another at an extreme phylo = 2.8 (bottom). While the phylo-based correction (orange) and the constant aperture correction (blue) give consistent results, the aperture correction derived from the UVIS1 encircled energy tables (based on Amp A) are $\sim$0.01 mag higher than the empirical values at nominal phylo and roughly equal to the Chip 1 values for the extreme phylo case (green dot-dashed line).
}}
\label{fig:tuccompare606}
\end{figure}

\begin{figure}[H]
\begin{center}
\includegraphics[width=\linewidth]{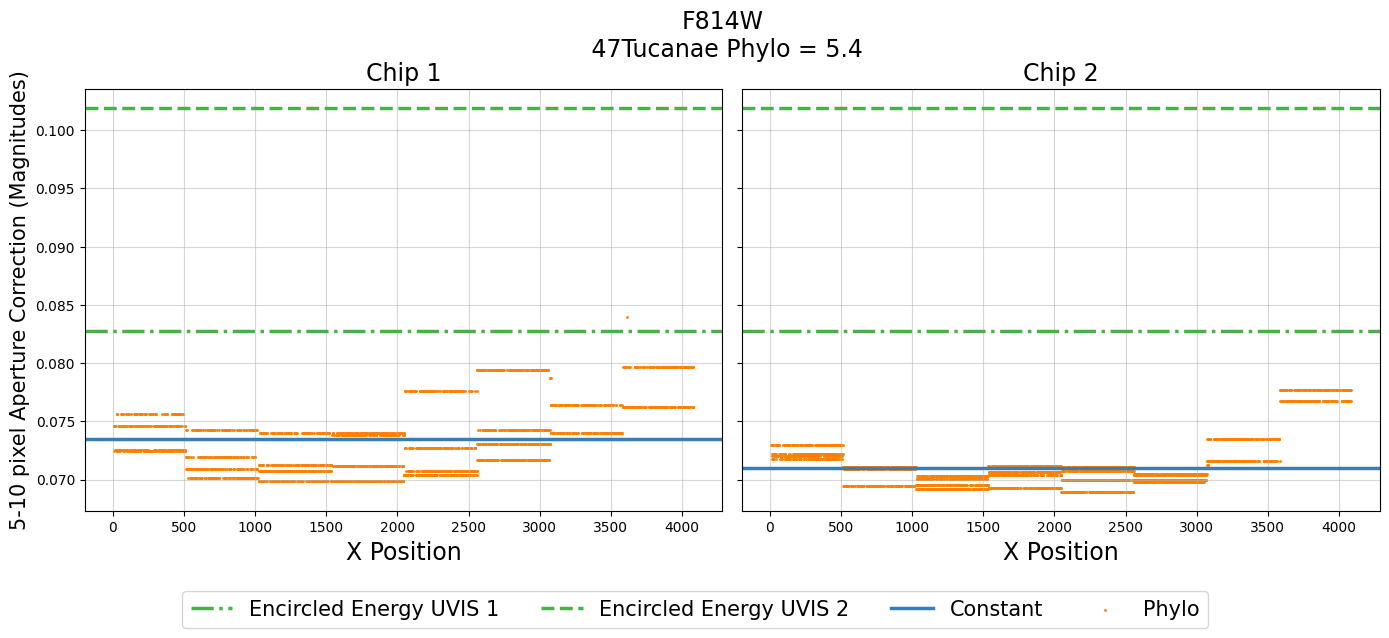}
\bigbreak
\includegraphics[width=\linewidth]{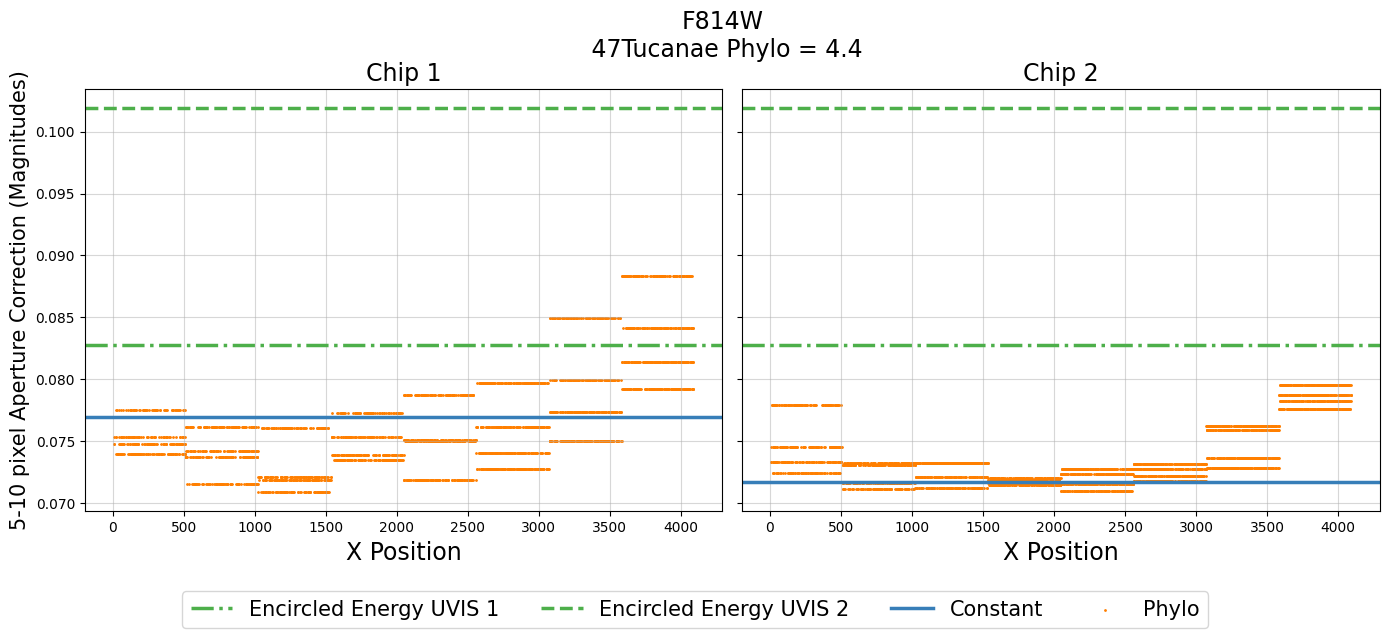}
\end{center}
\caption{\textit{\textit{Same as Figure \ref{fig:tuccompare438}, but for two F814W exposures of  47 Tucanae: one at nominal phylo = 5.4 (top) and another at phylo = 4.4 (bottom).  Note that observations in this filter have a narrower phylo range from 4.4 - 5.8, so the results are similar. In both cases, aperture corrections derived from the UVIS2 EE tables (green dashed lines) are significantly higher ($>$$\sim$0.02 mag) than the phylo-based values in orange and the constant correction in blue.}
}}
\label{fig:tuccompare814}
\end{figure}

In Figure \ref{fig:ocencompare438} and Figures \ref{fig:ocencompare606} and \ref{fig:ocencompare814} in the Appendix, we compare the 5--10 pixel constant, phylo-based, and EE aperture correction for two Omega Centauri images: one at nominal and another at an extreme phylo value for filters F438W, F606W, and F814W. Unlike the uncrowded region of 47 Tucanae shown in Figures \ref{fig:tuccompare438}, \ref{fig:tuccompare606}, and \ref{fig:tuccompare814}, the constant aperture correction estimated for every image is consistently overestimated compared to the phylo-based corrections, adding flux from contaminated stars to the measured photometry of the target star and systematically biasing the results.

\begin{figure}[H]
\begin{center}
\includegraphics[width=\linewidth]{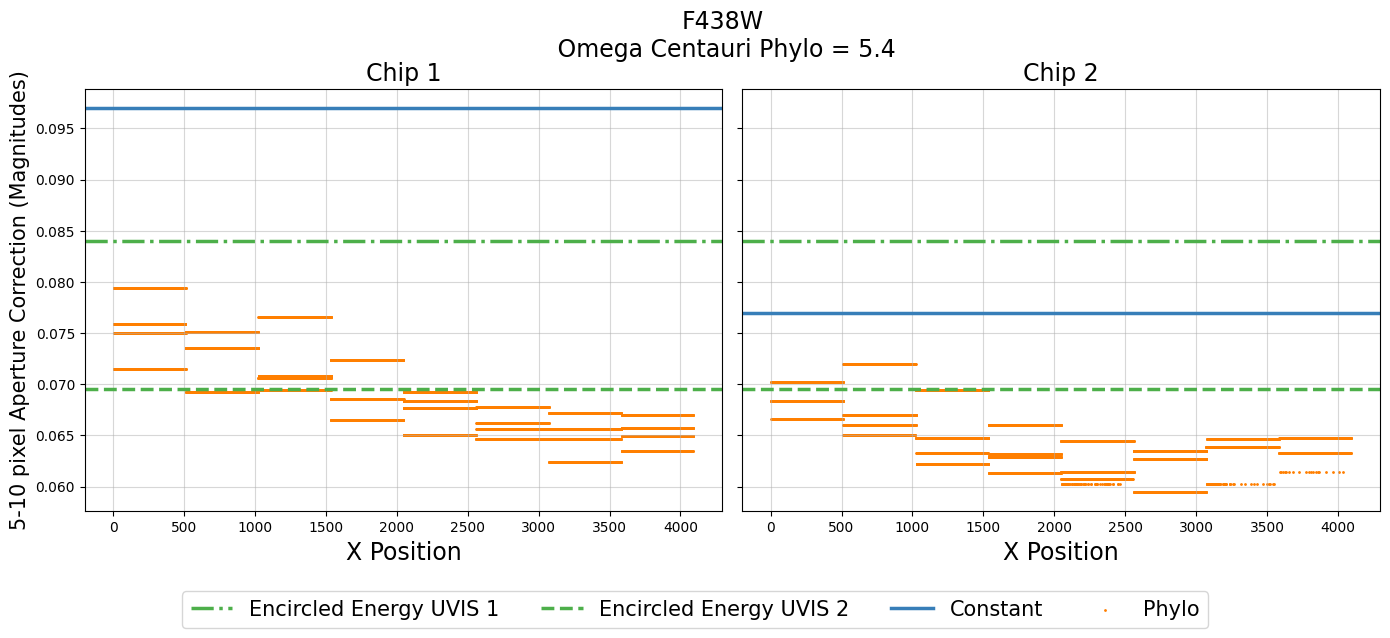}
\bigbreak
\includegraphics[width=\linewidth]{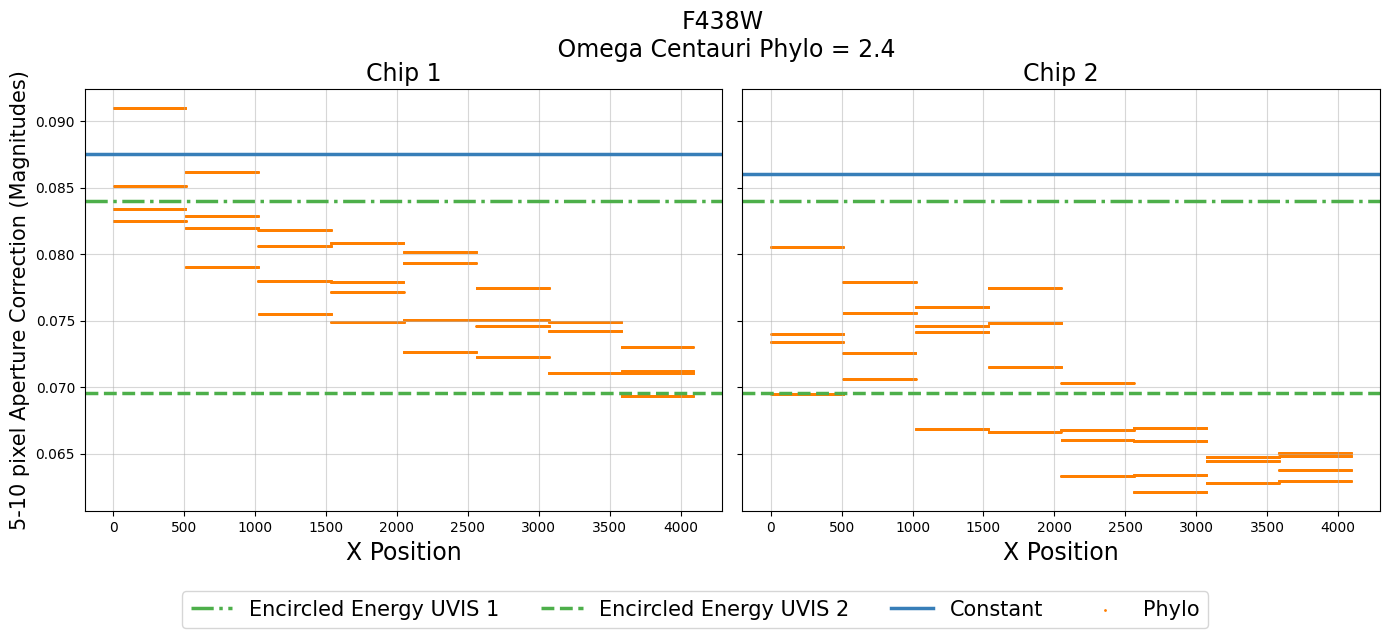}
\end{center}
\caption{\textit{F438W aperture correction (r = 5--10 pixels) versus X-position for two exposures in the crowded core of Omega Centauri: one at nominal phylo = 5.4 (top) and another at an extreme phylo value = 2.4 (bottom). For reference, aperture corrections derived from the encircled energy tables are overplotted in green. In both cases, the constant aperture correction (blue) is systematically larger than the phylo-based corrections (orange), highlighting the difficultly in accurately measuring the EE in crowded fields.}}
\label{fig:ocencompare438}
\end{figure}

\begin{figure}[H]
\begin{center}
\includegraphics[width=\textwidth]{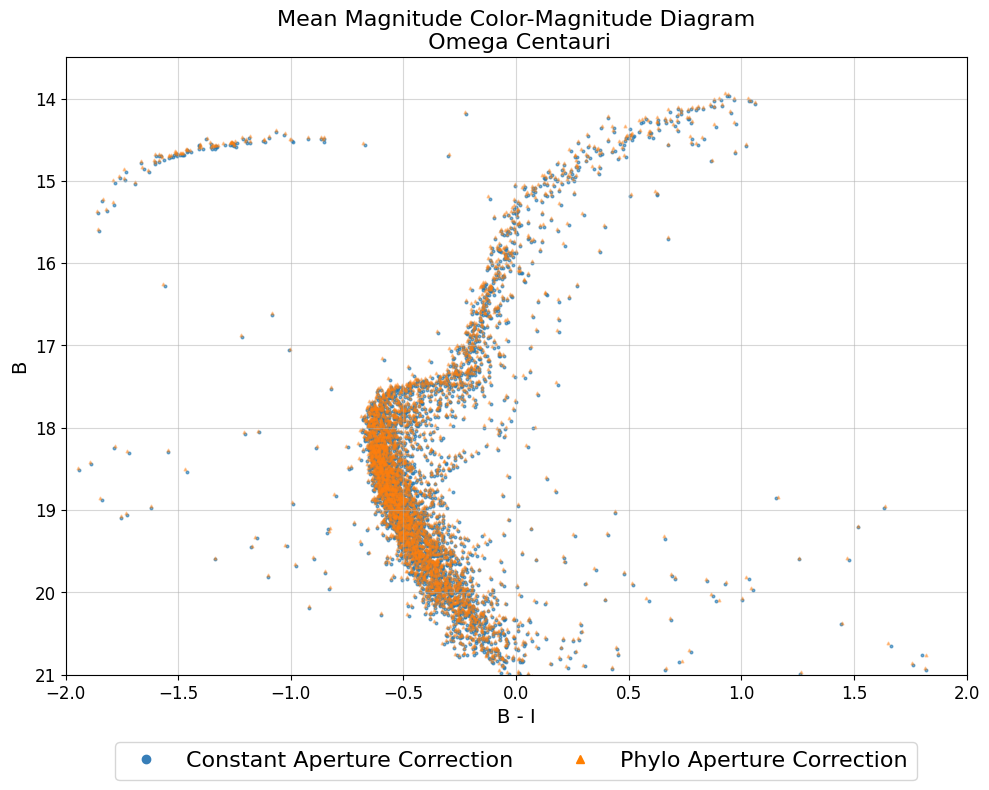}
\end{center}
\caption{\textit{Omega Centauri color-magnitude diagram derived from WFC3/UVIS B- and I- band (F438W and F814W) photometry in a 5 pixel aperture, corrected to 10 pixels using a constant (sigma-clipped mean) aperture correction per image (blue) versus our phylo-based aperture correction (orange). For both methods, the photometry is corrected from 10 pixels to infinite aperture using the EE tables. Colored points represent the mean magnitude of stars in a given filter for all Omega Centauri exposures from proposals 11911 and 12339. We measure a large shift of 0.024 magnitudes in the mean B-band magnitude between the two different methods and a negligible shift of 0.003 mag in the B-I color.}
}
\label{fig:omegacencmd}
\end{figure}

Figure \ref{fig:omegacencmd} shows a CMD derived from stars across all Omega Centauri exposures computed using B- and I- band (F438W and F814W) aperture photometry, corrected using two different methods: a constant (sigma-clipped mean) aperture correction per image and the new phylo-based aperture correction map.  The CMD derived using the new aperture corrections exhibits a large shift in the mean B-band magnitude of 0.024 mag and a negligible shift of 0.003 in the B--I color. This causes stars to appear fainter and slightly redder when using a constant aperture correction.

\subsection{Improvements in the Photometric Scatter}
\begin{figure}[H]
\begin{center}
\includegraphics[width=\textwidth]{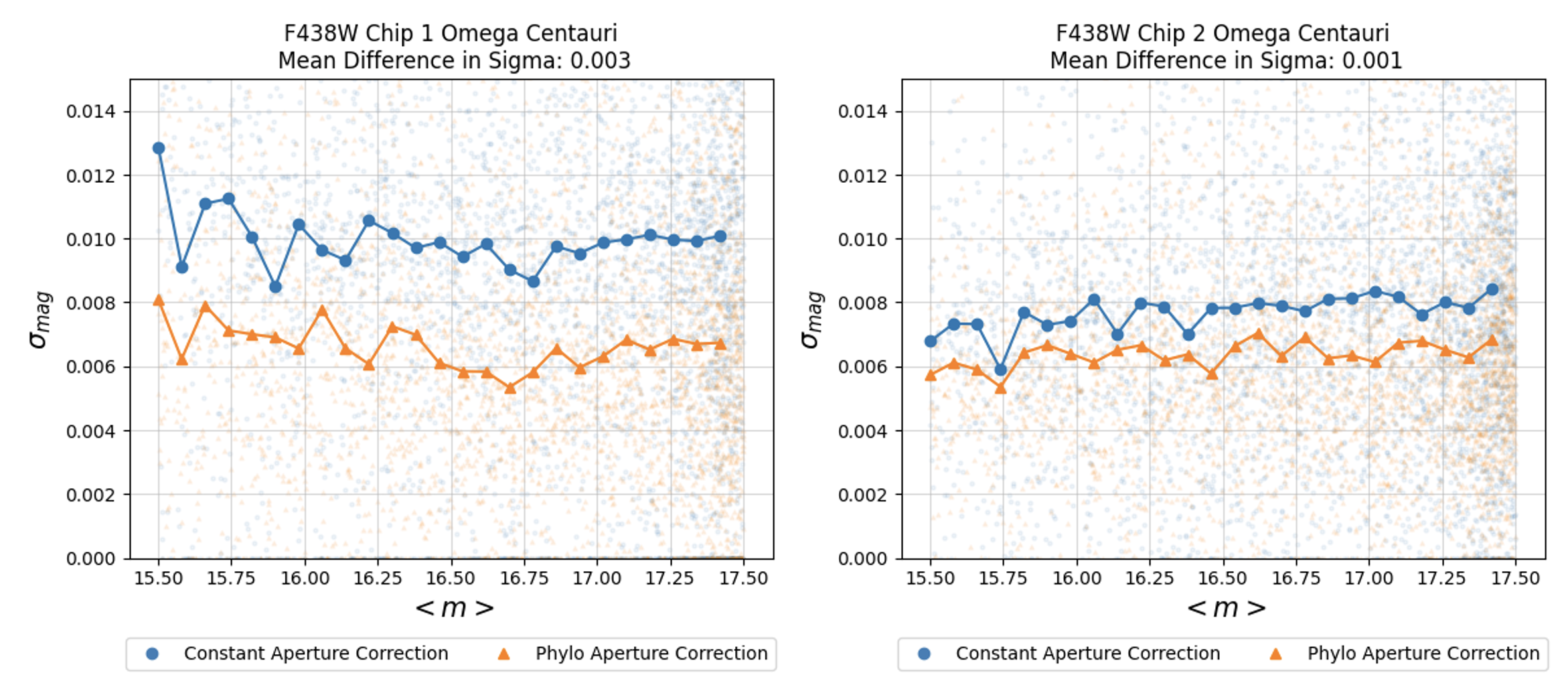}
\end{center}
\caption{\textit{Standard deviation versus mean magnitude of bright, unsaturated stars in F438W derived from dithered exposures of Omega Centuari. Blue points show the results after applying a constant (sigma-clipped mean) aperture correction per image, while orange points show the results after applying the new phylo-based aperture corrections. The median standard deviation using \textit{n} = 25 bins is over plotted. The phylo-based aperture correction improves the errors in the measurement of the mean magnitude by 0.003 mags in UVIS1 and 0.001 mags in UVIS2 compared to a constant aperture correction per CCD. This is consistent with the 2D phylo maps in Figures \ref{fig:2dmap438} and \ref{fig:apcorrscatter438} which show a consistently higher aperture correction in the Amp A corner in UVIS1 for almost all phylo values.}
}
\label{fig:sigmamag438}
\end{figure}

\begin{figure}[H]
\begin{center}
\includegraphics[width=\textwidth]{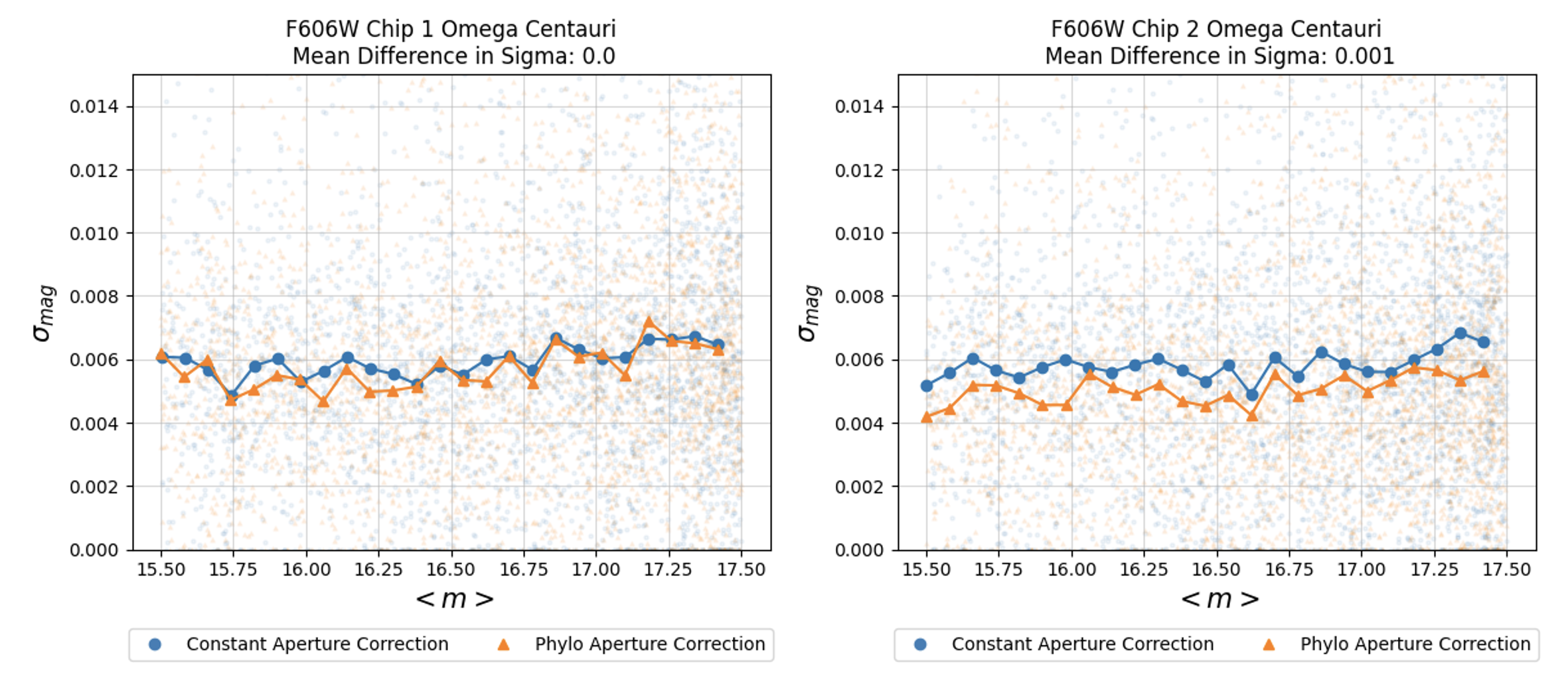}
\end{center}
\caption{\textit{Same as Figure \ref{fig:sigmamag438}, but for F606W. The phylo-based aperture correction improves the standard deviation in UVIS2 by 0.001 mag, but remains approximately the same in UVIS1, compared to a constant aperture correction per chip.}
}
\label{fig:sigmamag606}
\end{figure}

\begin{figure}[H]
\begin{center}
\includegraphics[width=\textwidth]{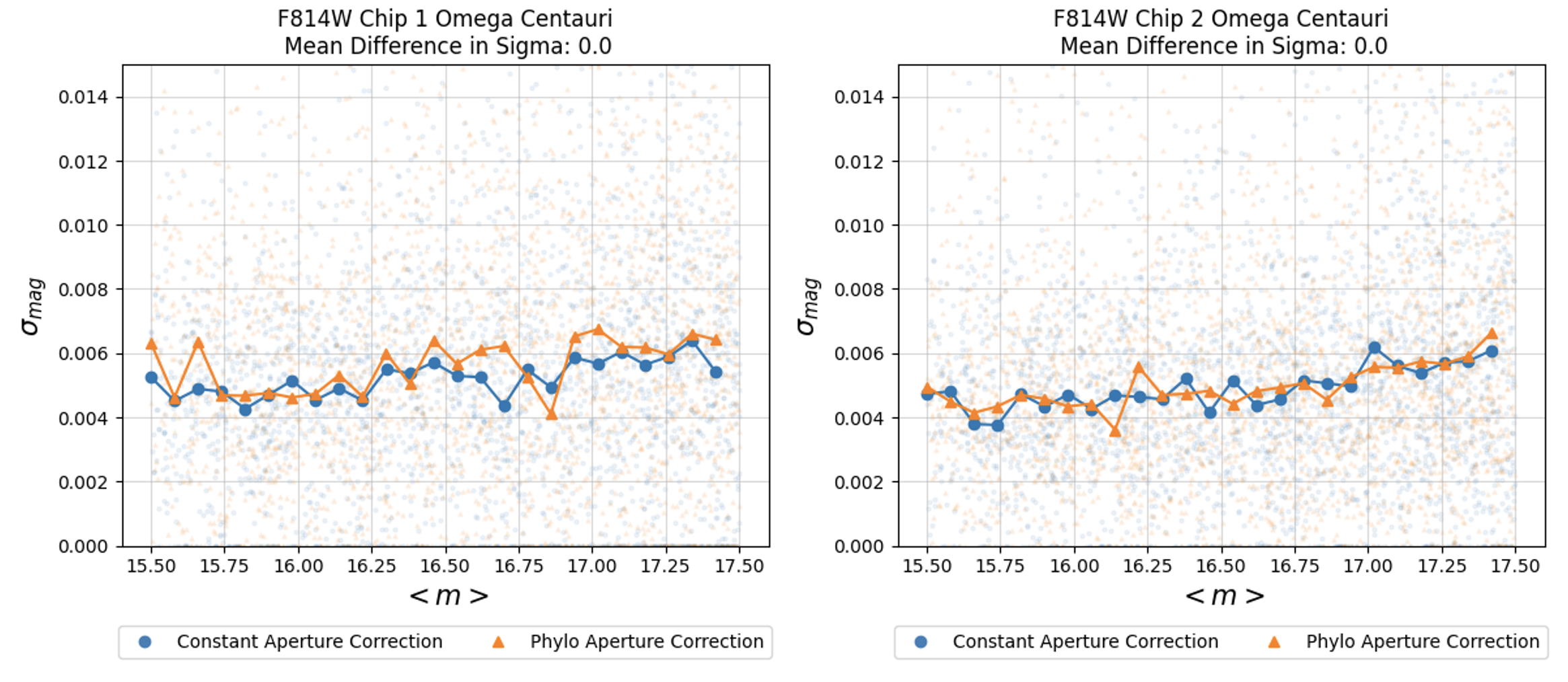}

\end{center}
\caption{\textit{Same as Figure \ref{fig:sigmamag438}, but for F814W. In this filter, there is no significant improvement in the standard deviation in either chip when comparing the phylo-based aperture correction versus a constant correction.}
}
\label{fig:sigmamag814}
\end{figure}

The phylo-based aperture correction also improves the scatter in the observed photometry and reduces the uncertainties in mean value of individual stars acquired in multiple dithered exposures. In Figures \ref{fig:sigmamag438}, \ref{fig:sigmamag606}, and \ref{fig:sigmamag814}, we plot the standard deviation versus the mean magnitude for bright, unsaturated stars measured across all dithered exposures of Omega Centauri in F438W, F606W, and F814W, respectively. A median-based binned statistic using \textit{n} = 25 bins is over-plotted to better visualize the improvement in the standard deviation when applying the phylo-based aperture corrections.
\bigbreak
The largest difference between the two methods is exhibited in F438W. The phylo-based aperture correction improves the errors in the measurement of the mean magnitude by 0.003 mag in UVIS1 and by 0.001 mag in UVIS2. This is consistent with results from the 2D phylo maps for F438W (illustrated in Figures \ref{fig:2dmap438} and \ref{fig:apcorrscatter438}), where bluer filters have a consistently higher aperture correction in the UVIS1 Amp A corner for almost all phylo values. In the F606W filter, we see slight improvements in UVIS2 only, where the standard deviation in the mean measurement is 0.001 magnitudes smaller when using the phylo-based aperture correction. There is however no noticeable improvement in the scatter measured in the F814W filter when comparing the photometry derived with constant and phylo-based aperture corrections.

\clearpage

\section{Conclusions}
We develop new 2D aperture correction maps in five filters (F275W, F336W, F438W, F606W, and F814W) based on the empirical focus level, or phylo, derived from measured PSFs. The new maps correct for temporal and spatial variations in aperture photometry between 5--10 pixels. Aperture corrections are smaller at nominal phylo levels (5--7) for all five filter maps. For bluer filters, the aperture correction in the Amp A corner is systematically higher than the rest of the FoV by $\sim$0.01 magnitudes for most phylo levels. Similarly, the aperture correction for redder filters is systematically higher than the rest of the FoV by $\sim$0.01 magnitudes.
\bigbreak
We test the improvements in aperture photometry of globular clusters using our phylo-based aperture correction maps, and we compare the results with those using a constant (sigma-clipped mean) aperture correction per exposure, measured from stars at radii of 5--10 pixels. Using dithered observations in three filters: F438W, F606W, and F814W, we test the improvement in photometry for an uncrowded region 6 arcminutes west of the core of 47 Tucanae and in a crowded region in the core of Omega Centauri. 
\bigbreak
For the uncrowded region, the constant aperture correction can be easily measured and compared with our 2D maps. For this target, our results can be summarized as follows:
\begin{enumerate}
    \item Aperture corrections derived from the published encircled energy tables are overestimated in most cases for F438W, F606W, and F814W, adding systematic uncertainties to the absolute photometry and color.  These tables may not reliable for small aperture photometry, where offsets exceeding $\sim$1\% are common when comparing to our phylo-based corrections or to a constant aperture correction per exposure.
    \item We create CMDs with a constant I-band and two different V-band exposures: one at nominal phylo and another at extreme phylo, and apply both a constant aperture correction per image and the phylo-based aperture correction. Using a constant aperture correction for two images at different phylo levels results in a shift of 0.016 magnitudes in the V-band photometry. This shift is reduced to 0.002 magnitudes when using the phylo-dependent correction. 
    \item The residual offset between the mean magnitude and the reference magnitude of the same star across different locations on the detector is relatively uniform throughout the FoV for bright, unsaturated stars when using a constant aperture correction versus a phylo-based aperture correction. The exception is in the Amp A corner in F438W and the Amp D corner in F814W where a $\sim$0.01 magnitude improvement is found when using the phylo-based aperture correction versus a constant in both cases.
\end{enumerate}
\bigbreak
In the crowded core of Omega Centauri, a constant aperture correction between 5--10 pixels is difficult to measure due to contamination effects. For this target, our validation results can be summarized as follows:
\begin{enumerate}
    \item We create mean magnitude color-magnitude diagrams using every B- and I-band exposure: one with a constant aperture correction per exposure and another with our phylo-based aperture correction. We find a shift of 0.024 magnitude in the B-band magnitude between measurements of stars using the two different methods and a negligible shift in the B-I color of 0.003 magnitudes. This causes stars to be fainter and slightly redder when using a constant correction, even after averaging multiple measurements of stars together in a crowded field.
    \item The phylo-based aperture correction improves the photometric scatter in F438W by 0.003 and 0.001 magnitudes in UVIS1 and UVIS2, respectively, for bright unsaturated stars in Omega Centauri compared to a constant aperture correction. The new maps improve the F606W scatter by 0.001 mag for UVIS2, but show no improvement for UVIS1. For F814W, no significant improvement in the scatter is seen for either chip.
\end{enumerate}

\section{Recommendation for Users}
\hypertarget{sec:rfu}{}
In crowded fields, photometry in small apertures can reduce systematic errors caused by neighbor contamination and background variations. The photometric keywords populated in the FITS image headers, however, are based on observations of CALSPEC standards measured at an ``infinite" aperture \parencite{2021Calamida}. 
EE corrections must therefore be applied to the observed photometry from finite apertures to compute the total brightness of a source.
\bigbreak
\textbf{While EE tables have been computed to large radii from deep images of the UVIS PSFs, application of these tables should be avoided for small apertures (r $<$ 10 pixels), where the EE varies with detector position and focus level.} 
\bigbreak
In this section, we highlight three methods to compute EE corrections for small apertures.

\subsection{Method 1: Use PSFs from science data (Recommended)}
EE corrections or aperture corrections should be measured using stars in each \texttt{FLC} science extension when possible, because sequential images within an HST orbit can span a range of focus levels. While drizzled (\texttt{DRC}) images may also be used, changes in the PSF at small radii will be averaged together in the combined frames.

To convert an aperture flux $F_r$ to a total magnitude $M$, two EE corrections are recommended: a correction from an aperture of radius $r$ to 10 pixels and a correction from 10 pixels to infinity, as shown in Equation \ref{meth1}:

\begin{equation}
\begin{split}
    M = -2.5 \log(F_r) + 2.5\log(\frac{EE_r}{EE_{r=10}}) + 
    2.5\log(\frac{EE_{r=10}}{EE_{r=inf}}) + ZP \\
    M =\,\,\,\,\,\,\,\,\,\,\, m_{r}\;\;\;\;\;\;\;+\;\;\;\,\,\;\;AP_{10}\;\;\;\;\;\;\;\;\;\;\;+\;\;\;\;\;\;AP_{inf}\;\;\;\;\;\;\;\;\;\;\;+ZP
    \\
\end{split}
    \label{meth1}
\end{equation}
where $m_{r}$ is the instrumental magnitude of the star at radius $r$, $AP_{10}$ is the aperture correction (mags) between radius $r$ and 10 pixels, $AP_{inf}$ is the aperture correction (mags) from 10 pixels to infinity from the EE tables, and $ZP$ is the magnitude zero point in the STMAG, ABMAG, or VEGAMAG photometric system.
\bigbreak
First, compute photometry at the smaller aperture and at the ‘standard’ aperture\footnote{For WFC3, the standard aperture is $\sim$0\farcs4 for both detectors (10 pixels for UVIS, 3 pixels for IR).}, beyond which the PSF is insensitive to changes in telescope focus, orbital breathing, or position on the detector ($AP_{10}$). Finally, use the EE tables\footnote{These tables were derived from deep images of isolated CALSPEC stars measured out to large radii (e.g. 151 pixels or 6\farcs0 for WFC3/UVIS).} to correct from 0\farcs4 to ‘infinite' aperture ($AP_{inf}$).\\
\\
\textbf{EXAMPLE:}

\begin{enumerate}[label=\textbf{Step \arabic*:},leftmargin=*]
  \item Measure the total flux $F_r$ of a star in a circular aperture of radius $r$ = 5 pixels observed in a drizzled \texttt{DRC} image in the center of UVIS2 for a 30 second exposure acquired in the F814W filter on June 26, 2009 (MJD 55008).  This date corresponds to the first WFC3 science observations, and is used as the ‘reference’ epoch for computing the inverse sensitivity (\texttt{PHOTFLAM}):
  \\ 
  \\
  $F_r = 1000~e^-/s$
  \\
  \\
  Alternatively, measure the total flux $f_r$ in the \texttt{FLC} image (in electrons), multiply by the pixel area map (PAM) and convert to count rate. (Note that flux is conserved between \texttt{FLC} and \texttt{DRC} images).
 \\
  \\
  $F_{r} = f_r * PAM / exptime$ \\
  $F_{r} = 29700 (e^-) * 1.01/30.0 (s) = 1000~e^-/s$
  \\
  \\
  where the pixel area correction at the center of UVIS2 is 1.01. PAMs are described in detail in \textcite{2010Kalirai} and are available for download on the \href{https://www.stsci.edu/hst/instrumentation/wfc3/data-analysis/pixel-area-maps}{WFC3 website}.

  \item Compute aperture photometry on bright, isolated stars in the image at $r=5$ and $r=10$ pixels, and then compute the mean flux ratio ($F_{r=5} /F_{r=10}$), which is equivalent to the EE ratio ($EE_{r=5}/EE_{r=10}$ ). We adopt the empirical mean aperture correction at phylo level 6 for F814W from Figure \ref{fig:apcorscatter_comb},
  e.g. apcor=0.069 mag, which gives a mean flux ratio:
\\
\\
$F_{r=5} /F_{r=10} = 10^{(0.069/-2.5)} = 0.938$
\\
\item Use the Encircled Energy tables to find the EE at 10 pixels (0\farcs4). UVIS EE tables are described in \textcite{2021Calamida} and CSV tables are available for download on the \href{https://www.stsci.edu/hst/instrumentation/wfc3/data-analysis/photometric-calibration/uvis-encircled-energy}{WFC3 website}. For F814W, the EE for both UVIS1 and UVIS2 is 0.902 at 10 pixels.
\item Convert from observed count rate to total magnitude using Equation \ref{meth1}:
\\
\\
$M = -2.5 log(1000) + 2.5 log(0.938) + 2.5 log(0.902) + ZP$
\\
\\
where $ZP$ in the STMAG system is related to \texttt{PHOTFLAM}, the inverse sensitivity at infinite aperture (erg s$^{-1}$ cm$^{-2}$ Å$^{-1}$ per e$^{-}$ s$^{-1}$) on the date of observation and \texttt{PHOTZP} (the STMAG zeropoint). For F814W, the zeropoint in the equation for total magnitude is computed from keywords in the image header corresponding to MJD 55008:
\\
\\
$ZP = - 2.5 log(\texttt{PHOTFLAM}) + \texttt{PHOTZP} = $$- 2.5 log(1.498 * 10^{-19}) -21.1 = 25.961$
\\
\\
This gives a total magnitude:
\\
\\
$M = -7.500 -0.069 - 0.112 +25.961 $= \textbf{18.280 mag}
\end{enumerate}

\subsection{Method 2: Use PSF cutouts from MAST}
The PSF search tool in MAST can be used to download stellar cutouts extracted from archival data at similar detector positions and focus levels (e.g. see the \href{https://www.stsci.edu/hst/instrumentation/focus-and-pointing/focus/hst-focus-model}{HST focus model page}). The model reports the focus in units of microns of despace, which is related to the change in spacing between the primary and secondary mirror of HST \parencite{2024Dressel}. While the focus is related to the phylo level computed by \texttt{hst1pass}  \parencite{18Anderson}, a one-to-one correspondence between the two has not yet been computed. Instead, users can query the predicted focus level for their exposures and look for archival exposures at a similar focus level using this model. The focus model is updated periodically, though is not guaranteed to be in sync with phylo measurements (which are measured in each image) as the errors in the model compound over time.
\bigbreak
\href{https://www.stsci.edu/hst/instrumentation/wfc3/data-analysis/psf/psf-search}{WFC3 Observed PSFs} can be accessed on the \href{https://mast.stsci.edu/portal/Mashup/Clients/Mast/Portal.html}{MAST Portal interface} by choosing ``WFC3 PSF" under ``Select a Collection". Cutouts near the center of the detector where the PSF is less sensitive to focus are recommended. For details, see \textcite{Dauphin2021}.
\bigbreak

Here, the workflow is identical to Method 1, except the mean flux ratio ($F_{r=5} /F_{r=10}$) in Step 2 is measured in each of the PSF cutouts and then averaged to estimate the EE ratio ($EE_{r=5} /EE_{r=10}$ ) at a given focus level and position on the detector.

\subsection{Method 3: Use the EE tables (least accurate)}

In this method, we compute aperture corrections from the published EE tables and compare the total magnitude derived from Equation \ref{eq:EEeq} with the results from Method 1.  

\begin{equation}
M_{UVIS2} = -2.5log(F_{r}) + 2.5log(\frac{\text{UVIS2}\_EE_{r}}{EE_{r=inf}}) + ZP		
\label{eq:EEeq}
\end{equation}

For a star with an observed flux $F_{r=5} = 1000~e^-/s$, the UVIS2 EE in F814W at a radius r=5 pixels is 0.821. The total magnitude is calculated as:
\\
\\
$M_{UVIS2} = -2.5log(1000)+ 2.5log(0.821/1.000) + ZP $
\\
$M_{UVIS2} = -7.500 - 0.214 + 25.961=$  \textbf{18.247 mag}
\\
\\
Note that the difference from the true magnitude of the star is:
\\
\\
$dM_{UVIS2}= M - M_{UVIS2} = 18.280 - 18.247 = $ \textbf{\textit{0.033 mag}}
\\
\\
Alternatively, the UVIS1 EE at r=5 pixels is 0.836.  This gives a total magnitude:
\\
\\
$M_{UVIS1} = -2.5 log(F_{r}) + 2.5 log(\frac{UVIS1\_EE_{r}}{EE_{r=inf}}) + ZP$
\\
$M_{UVIS1} = -2.5 log(10000)+ 2.5 log(0.836/1.000) +ZP$
\\
$M_{UVIS1}=  -7.500  -0.195 + 25.961 =$ \textbf{18.266 mag}
\\
\\
While the UVIS2 EE is typically a better estimate of the EE over the detector FoV, we find that the UVIS1 EE table gives a slightly more accurate result for F814W:
\\
\\
$dM_{UVIS1} = M - M_{UVIS1} = 18.280 - 18.266 = $ \textbf{\textit{0.014 mag}}
\\
\\
This is consistent with results in Figure \ref{fig:apcorscatter_comb}, where the F814W UVIS2 EE tables give aperture corrections that are $\sim$0.03 mag larger than the corrections from the phylo-dependent values between 5--10 pixels.  A similar offset is observed for the F275W UVIS2 EE table. The UVIS2 EE tables for F336W, F438W, and F606W are within $\sim$0.005 mag of the average phylo-based correction for Amp C. This can cause a horizontal shift in the color-magnitude diagram, for example, when plotting B vs B-I or V vs V-I, where only the F814W filter is offset from the phylo-based values, as seen in Figures \ref{fig:47tuccmd} and \ref{fig:omegacencmd}.
\bigbreak
The UVIS1 EE tables are within $\sim$0.01 mag of the average phylo-based correction for Amp A, but these values are systematically higher than the mean correction over the FoV and are recommended only for sources in the upper-left corner of the detector.  
\bigbreak
\textbf{To reduce errors in absolute photometry to better than $\sim1\%$, aperture corrections should be measured from isolated stars in the science data or from PSF cutouts at a similar detector position and focus.}


\section{Acknowledgements}
\hypertarget{sec:acknowledgements}{}
The authors would like to thank Mitchell Revalski, Joel Green, and Sylvia Baggett for their thorough review of this report. We thank Jay Anderson for his work developing the focus-diverse PSF models and the \texttt{hst1pass} software, as it was the basis of the methods to compute phylo levels for a large set of UVIS archival images. We are grateful to the rest of the WFC3 photometry team for their in-depth discussions and assistance for writing this report. In particular, we thank Anne O'Connor for her work testing potential improvements to the current encircled energy tables at small apertures. 
\printbibliography

\newpage
\section{Appendix}
\hypertarget{sec:appendix}{}

\subsection{Comparing Aperture Correction Methods in 47 Tucanae}

\newpage
\subsection{Comparing Aperture Correction Methods in Omega Centauri}
\begin{figure}[H]
\begin{center}
\includegraphics[width=\linewidth]{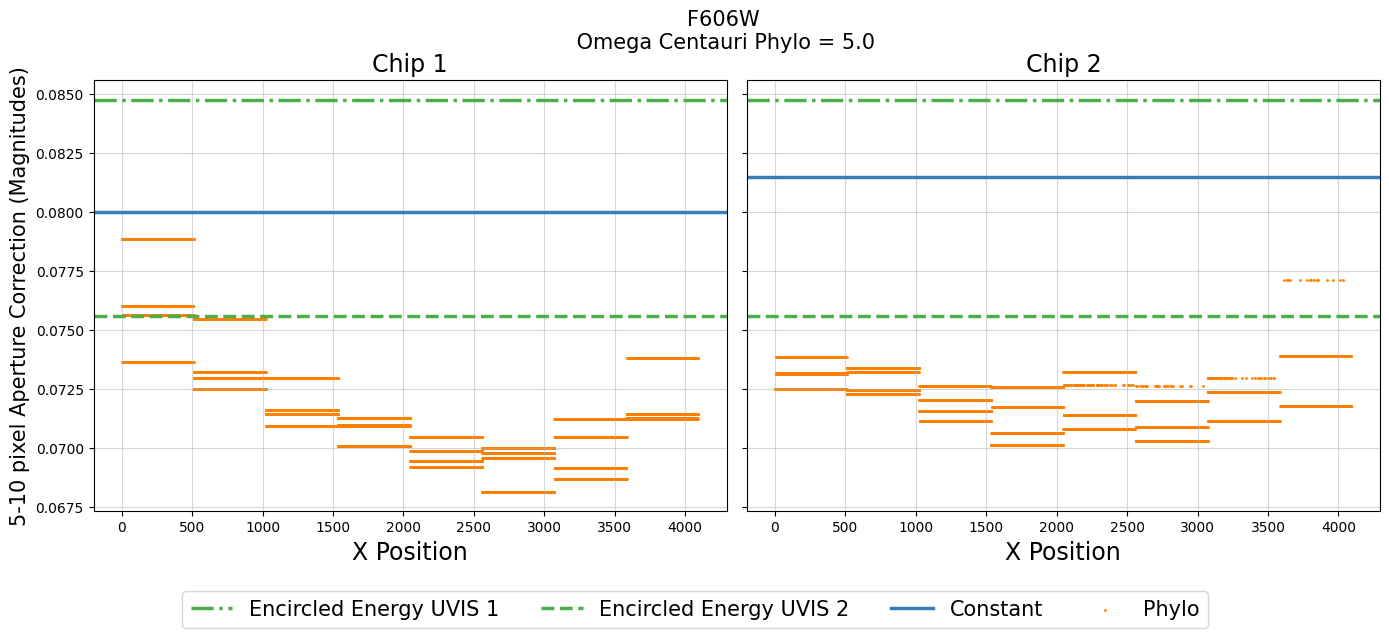}
\bigbreak
\includegraphics[width=\linewidth]{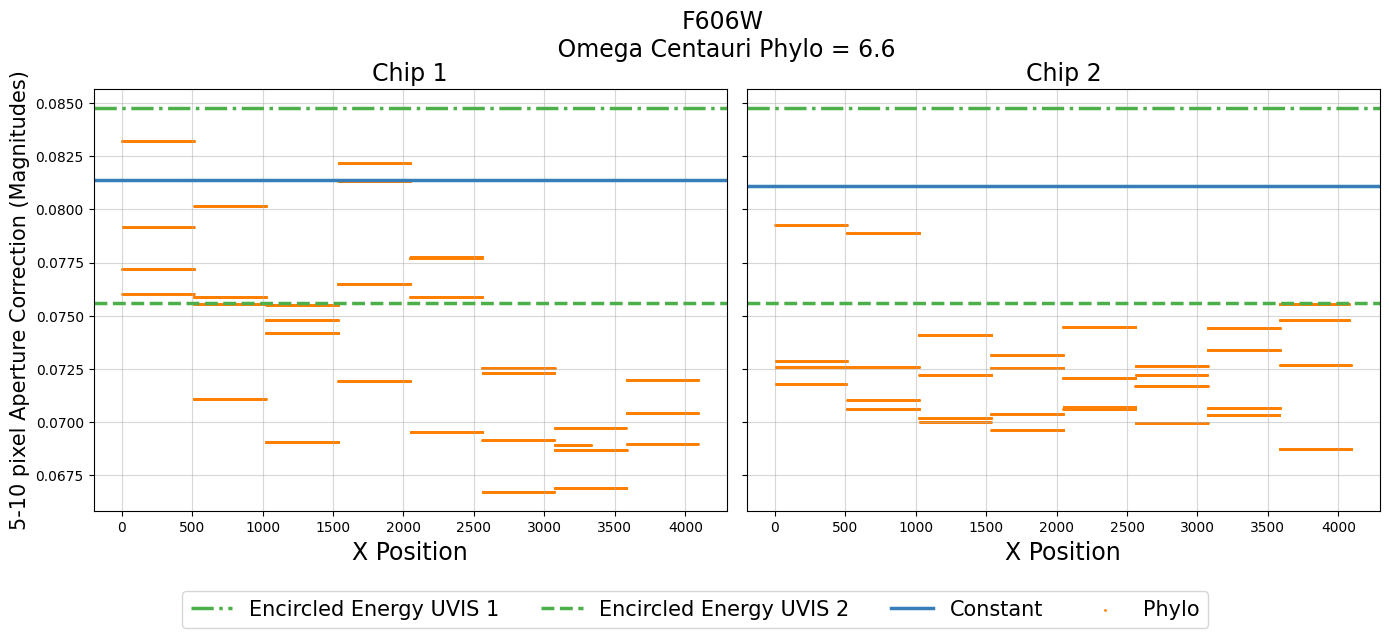}
\end{center}
\caption{\textit{Same as Figure \ref{fig:ocencompare438}, but for two F606W exposures in the crowded core of Omega Centauri: one at nominal phylo = 5.0 (top) and another at an extreme phylo = 6.6 (bottom). Note that observations in this filter have a narrower phylo  range from 4.2 - 6.6, so the results are similar.}
}
\label{fig:ocencompare606}
\end{figure}

\begin{figure}[H]
\begin{center}
\includegraphics[width=\linewidth]{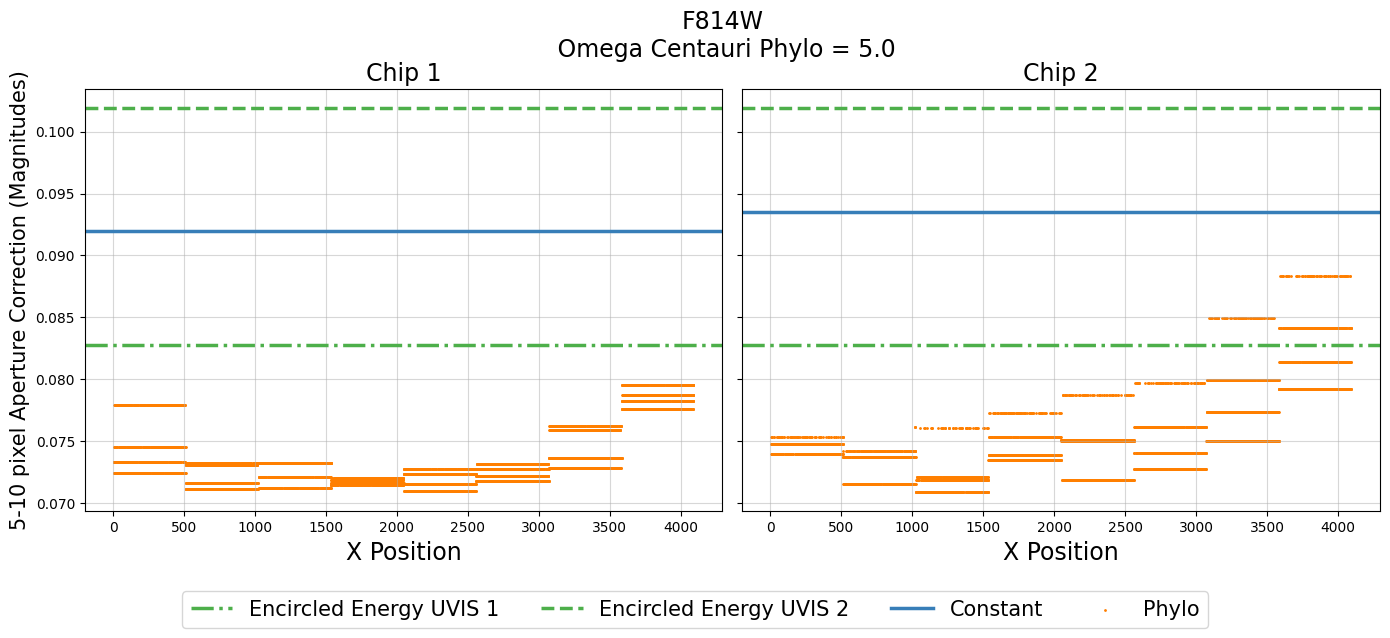}
\bigbreak
\includegraphics[width=\linewidth]{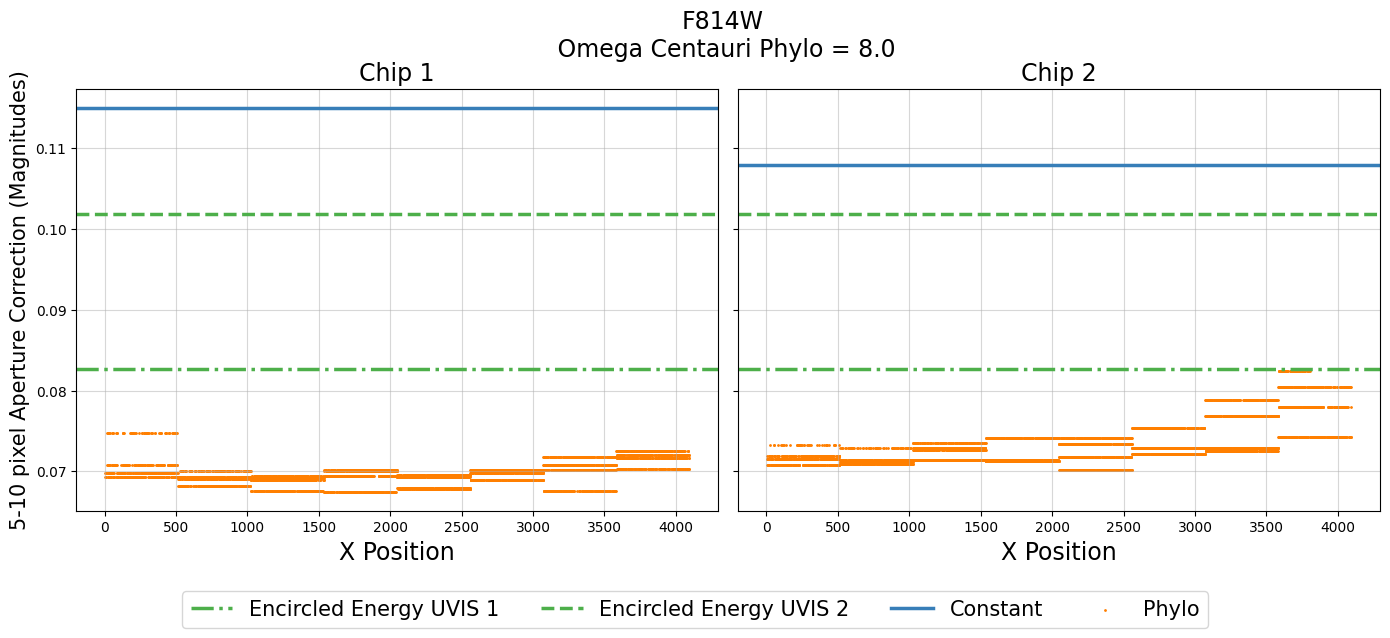}
\end{center}
\caption{\textit{Same as Figure \ref{fig:ocencompare814}, but for two F814W exposures in the crowded core of Omega Centauri: one at nominal phylo = 5.0 (top) and another at an extreme phylo = 8.0 (bottom). The constant aperture correction in blue is systematically higher from the phylo-dependent correction in orange. Similarly, the aperture corrections derived from the UVIS2 EE tables (green dashed lines) are consistently too large for this filter. }
}
\label{fig:ocencompare814}
\end{figure}

\newpage
\subsection{Tables}
\begin{table}[h!]
    \centering
    \captionsetup{justification=centering} 
    \caption{\textit{Rootname, filter, proposal ID, and phylo value measured for 47 Tucanae exposures used to test the new phylo-based aperture corrections. These data are from programs 11452 and 12379, which were originally used to test the UVIS flat field uniformity \parencite{Sabbi2009} and to measure the CTE, respectively.}}
    \label{tab:47tucdata}
\end{table}

\vspace{-1cm}
\begin{paracol}{2}
\begin{table}[H]
\begin{tabular}{c c c c}
\toprule
Rootname & Filter & Proposal ID & Phylo \\
\midrule
iaby01kcq & F438W & 11452 & 5.8 \\
iaby01lgq & F438W & 11452 & 7.4 \\
iaby01liq & F438W & 11452 & 7.6 \\
iaby01lkq & F438W & 11452 & 7.8 \\
iaby01lmq & F438W & 11452 & 7.8 \\
iaby01loq & F438W & 11452 & 7.6 \\
iaby01lqq & F438W & 11452 & 6.0 \\
iaby01lsq & F438W & 11452 & 6.0 \\
iaby01luq & F438W & 11452 & 7.4 \\
\hline
iaby01khq & F606W & 11452 & 3.8 \\
iaby01klq & F606W & 11452 & 4.6 \\
iaby01kmq & F606W & 11452 & 4.6 \\
iaby01koq & F606W & 11452 & 4.0 \\
iaby01kqq & F606W & 11452 & 3.8 \\
iaby01ksq & F606W & 11452 & 4.6 \\
iaby01kuq & F606W & 11452 & 4.6 \\
iaby01kwq & F606W & 11452 & 4.8 \\
iaby01kyq & F606W & 11452 & 4.2 \\
ibnh02c5q & F606W & 12379 & 4.8 \\
ibnh07nkq & F606W & 12379 & 4.4 \\
ibnh07nnq & F606W & 12379 & 3.8 \\
ibnh07nrq & F606W & 12379 & 3.6 \\
ibnh07nvq & F606W & 12379 & 5.0 \\
ibnh07nzq & F606W & 12379 & 5.0 \\
ibnh07o3q & F606W & 12379 & 4.0 \\
ibnh07o7q & F606W & 12379 & 3.8 \\
ibnh07obq & F606W & 12379 & 4.6 \\
ibnh07ofq & F606W & 12379 & 5.0 \\
ibnh07ojq & F606W & 12379 & 4.2 \\
ibnh08hlq & F606W & 12379 & 4.4 \\
\bottomrule
\end{tabular}
\end{table}
\switchcolumn


\begin{table}[H]
\begin{tabular}{c c c c}
\toprule
Rootname & Filter & Proposal ID & Phylo \\
\midrule
ibnh11beq & F606W & 12379 & 4.6 \\
ibnh11bgq & F606W & 12379 & 4.8 \\
ibnh11biq & F606W & 12379 & 5.0 \\
ibnh11bkq & F606W & 12379 & 5.8 \\
ibnh11bmq & F606W & 12379 & 5.0 \\
ibnh11byq & F606W & 12379 & 5.2 \\
ibnh13guq & F606W & 12379 & 4.2 \\
ibnh13gwq & F606W & 12379 & 4.6 \\
ibnh13gyq & F606W & 12379 & 4.6 \\
ibnh13h0q & F606W & 12379 & 4.6 \\
ibnh13h2q & F606W & 12379 & 5.4 \\
ibnh13heq & F606W & 12379 & 5.0 \\
ibnh14x5q & F606W & 12379 & 2.8 \\
ibnh14x7q & F606W & 12379 & 2.8 \\
ibnh14xgq & F606W & 12379 & 4.6 \\
ibnh14xiq & F606W & 12379 & 3.2 \\
ibwb02ekq & F606W & 12692 & 4.4 \\
ibwb05dnq & F606W & 12692 & 3.6 \\
ibwb08a4q & F606W & 12692 & 3.4 \\
ibwb08a6q & F606W & 12692 & 3.2 \\
\hline
iaby01lcq & F814W & 11452 & 4.6 \\
iaby01leq & F814W & 11452 & 4.6 \\
iaby01kgq & F814W & 11452 & 4.6 \\
iaby01l6q & F814W & 11452 & 4.4 \\
iaby01l0q & F814W & 11452 & 5.0 \\
iaby01l2q & F814W & 11452 & 5.0 \\
iaby01l8q & F814W & 11452 & 5.4 \\
iaby01laq & F814W & 11452 & 5.6 \\
iaby01l4q & F814W & 11452 & 4.6 \\
\bottomrule
\end{tabular}
\end{table}
\end{paracol}

\newpage
\begin{table}[h!]
    \centering
    \captionsetup{justification=centering} 
    \caption{\textit{Rootname, filter, proposal ID, and phylo value for Omega Centauri exposures used to test the new phylo-based aperture corrections. These data are from programs 11911 and 12339 which were originally used to derive in-flight corrections to the UVIS flat fields \parencite{2013mack}.}}
    \label{tab:ocendata}
\end{table}
\vspace{-1cm}
\begin{paracol}{2}
\begin{table}[H]

\centering
\begin{tabular}{c c c c}
\toprule
Rootname & Filter & Proposal ID & Phylo \\
\midrule
ibc301qxq & F438W & 11911 & 4.8 \\
ibc301r7q & F438W & 11911 & 4.2 \\
ibc301r9q & F438W & 11911 & 4.6 \\
ibc301riq & F438W & 11911 & 4.2 \\
ibc301rpq & F438W & 11911 & 2.4 \\
ibc301rrq & F438W & 11911 & 3.0 \\
ibc304w5q & F438W & 11911 & 3.6 \\
ibc304wgq & F438W & 11911 & 4.0 \\
ibc304wiq & F438W & 11911 & 4.6 \\
ibc304wkq & F438W & 11911 & 7.0 \\
ibc304wmq & F438W & 11911 & 7.2 \\
ibc304wyq & F438W & 11911 & 3.0 \\
ibc304x0q & F438W & 11911 & 4.0 \\
ibc307tlq & F438W & 11911 & 5.4 \\
ibc308xuq & F438W & 11911 & 7.4 \\
ibc308xwq & F438W & 11911 & 6.4 \\
ibc308xyq & F438W & 11911 & 6.0 \\
ibc308y0q & F438W & 11911 & 6.0 \\
ibc308y2q & F438W & 11911 & 6.0 \\
ibc308y7q & F438W & 11911 & 6.2 \\
ibc308y9q & F438W & 11911 & 6.4 \\
ibc308ybq & F438W & 11911 & 6.0 \\
ibla01f3q & F438W & 12339 & 8.0 \\
ibla02srq & F438W & 12339 & 2.2 \\
ibla02stq & F438W & 12339 & 3.6 \\
ibla02svq & F438W & 12339 & 4.2 \\
ibla02sxq & F438W & 12339 & 7.0 \\
ibla02szq & F438W & 12339 & 7.0 \\
ibla02t7q & F438W & 12339 & 4.0 \\
ibla02t9q & F438W & 12339 & 7.0 \\
ibla02tbq & F438W & 12339 & 7.0 \\
\hline
ibc302ivq & F606W & 11911 & 6.4 \\
ibc302j0q & F606W & 11911 & 5.4 \\
ibc302j7q & F606W & 11911 & 5.8 \\
ibc302jcq & F606W & 11911 & 6.6 \\

\bottomrule
\end{tabular}
\end{table}
\switchcolumn
\begin{table}[H]
\centering
\begin{tabular}{c c c c}
\toprule
Rootname & Filter & Proposal ID & Phylo \\
\midrule
ibc303n1q & F606W & 11911 & 6.6 \\
ibc303n9q & F606W & 11911 & 5.0 \\
ibc304v3q & F606W & 11911 & 6.2 \\
ibc306q7q & F606W & 11911 & 4.2 \\
ibc306qqq & F606W & 11911 & 4.6 \\
ibc306qsq & F606W & 11911 & 4.4 \\
ibc307qyq & F606W & 11911 & 4.2 \\
ibc307raq & F606W & 11911 & 5.0 \\
ibc307rgq & F606W & 11911 & 5.0 \\
ibc307rpq & F606W & 11911 & 4.2 \\
ibc307sjq & F606W & 11911 & 5.0 \\
ibc307soq & F606W & 11911 & 5.0 \\
ibc307sqq & F606W & 11911 & 5.2 \\
ibc307stq & F606W & 11911 & 4.2 \\
ibc307syq & F606W & 11911 & 5.2 \\
ibla01drq & F606W & 12339 & 3.6 \\
ibla01dwq & F606W & 12339 & 5.4 \\
ibla01dyq & F606W & 12339 & 4.2 \\
ibla01e1q & F606W & 12339 & 3.6 \\
ibla01e6q & F606W & 12339 & 3.2 \\
ibla01e8q & F606W & 12339 & 5.2 \\
ibla01ebq & F606W & 12339 & 4.2 \\
ibla01egq & F606W & 12339 & 3.6 \\
\hline
ibc301qzq & F814W & 11911 & 6.4 \\
ibc307tnq & F814W & 11911 & 7.4 \\
ibc308ydq & F814W & 11911 & 7.2 \\
ibc307t0q & F814W & 11911 & 6.0 \\
ibc302ijq & F814W & 11911 & 7.2 \\
ibc302hsq & F814W & 11911 & 6.2 \\
ibc302ioq & F814W & 11911 & 7.0 \\
ibc307t3q & F814W & 11911 & 6.8 \\
ibla02tdq & F814W & 12339 & 7.8 \\
ibc307t8q & F814W & 11911 & 8.0 \\
ibla01esq & F814W & 12339 & 6.0 \\
ibla01elq & F814W & 12339 & 4.6 \\
\bottomrule
\end{tabular}
\end{table}
\end{paracol}
\newpage
\begin{table}[h!]
    \centering
    \captionsetup{justification=centering} 
    \caption*{\textit{Table \ref{tab:ocendata} continued}.}
\end{table}
\vspace{-1cm}
\begin{paracol}{2}
\begin{table}[H]
\centering
\begin{tabular}{c c c c}
\toprule
Rootname & Filter & Proposal ID & Phylo \\
\midrule
ibc308xtq & F814W & 11911 & 6.0 \\
ibla02t3q & F814W & 12339 & 6.2 \\
ibc305y4q & F814W & 11911 & 6.0 \\
ibla01eiq & F814W & 12339 & 5.2 \\
ibla02sqq & F814W & 12339 & 5.0 \\
ibc305xuq & F814W & 11911 & 4.4 \\
ibc308y6q & F814W & 11911 & 5.6 \\
ibla02t1q & F814W & 12339 & 8.0 \\
\bottomrule
\end{tabular}
\end{table}
\switchcolumn
\begin{table}[H]
\centering
\begin{tabular}{c c c c}
\toprule
Rootname & Filter & Proposal ID & Phylo \\
\midrule
ibla02t4q & F814W & 12339 & 4.8 \\
ibc308y4q & F814W & 11911 & 4.6 \\
ibc305xkq & F814W & 11911 & 4.8 \\
ibla01eqq & F814W & 12339 & 8.0 \\
ibc307taq & F814W & 11911 & 4.6 \\
ibc302i7q & F814W & 11911 & 8.0 \\
ibc304w8q & F814W & 11911 & 8.0 \\
ibc302icq & F814W & 11911 & 4.8 \\
\bottomrule
\end{tabular}
\end{table}
\end{paracol}

\end{document}